\documentclass[final,3p,times]{elsarticle}
\usepackage[latin9]{inputenc}
\usepackage{color}
\usepackage{array}
\usepackage{bm}
\usepackage{multirow}
\usepackage{amsmath}
\usepackage{graphicx}
\usepackage[unicode=true, bookmarks=false, breaklinks=false,pdfborder={0 0 1},backref=section,colorlinks=false]{hyperref}

\makeatletter

\providecommand{\tabularnewline}{\\}
\newcommand{\lyxdot}{.}

\usepackage{lineno}
\modulolinenumbers[5]

\journal{Probabilistic Engineering Mechanics}

\makeatother

\begin{document}
\begin{frontmatter}

\title{A polynomial dimensional decomposition framework based on topology
derivatives for stochastic topology sensitivity analysis of high-dimensional
complex systems and a type of benchmark problems}


\author{Xuchun Ren}

\cortext[correspondingauthor]{Corresponding author}

\ead{NHABEL@gmail.com; xuchun.ren777@gmail.com}

\address[correspondingauthor]{Mechanical Engineering Department, Georgia Southern University,
Statesboro, GA, USA}
\begin{abstract}
In this paper, a new computational framework based on the topology
derivative concept is presented for evaluating stochastic topological
sensitivities of complex systems. The proposed framework, designed
for dealing with high dimensional random inputs, dovetails a polynomial
dimensional decomposition (PDD) of multivariate stochastic response
functions and deterministic topology derivatives. On one hand, it
provides analytical expressions to calculate topology sensitivities
of the first three stochastic moments which are often required in
robust topology optimization (RTO). On another hand, it offers embedded
Monte Carlo Simulation (MCS) and finite difference formulations to
estimate topology sensitivities of failure probability for reliability-based
topology optimization (RBTO). For both cases, the quantification of
uncertainties and their topology sensitivities are determined concurrently
from a single stochastic analysis. Moreover, an original example of
two random variables is developed for the first time to obtain analytical
solutions for topology sensitivity of moments and failure probability.
Another 53-dimension example is constructed for analytical solutions
of topology sensitivity of moments and semi-analytical solutions of
topology sensitivity of failure probabilities in order to verify the
accuracy and efficiency of the proposed method for high-dimensional
scenarios. Those examples are new and make it possible for researchers
to benchmark stochastic topology sensitivities of existing or new
algorithms. In addition, it is unveiled that under certain conditions
the proposed method achieves better accuracies for stochastic topology
sensitivities than for the stochastic quantities themselves. 
\end{abstract}
\end{frontmatter}

%\linenumbers

\section{Introduction \label{sec1}}

With the rise of additive manufacturing, topology optimization becomes
a popular design methodology to determine the optimal distribution
of materials in complex engineering structures\citep{yunkang1998new,wang2003level,wang2004color,liu2016multi}.
Inevitable uncertainties in the additive manufacturing process and
operating environment often undermine the performance of such topology
designs. Classical deterministic design approaches often lead to unknowingly
risky designs due to the underestimation of uncertainties, or inefficient
and conservative designs that overcompensate for uncertainties. In
the past decade, robust topology optimization (RTO) and reliability-based
topology optimization (RBTO) are increasingly adopted as an enabling
technology for topology design subject to uncertainty in aerospace,
automotive, civil engineering, and additive manufacturing \citep{chen2010level,chen2011new,asadpoure2011robust,guo2013robust,guo2015multi,zhang2016robust,jiang2017parametric}.
The former seeks for insensitive topology design via minimizing the
propagation of input uncertainty, whereas the latter delivers reliable
topology design by introducing probabilistic characterizations of
response functions into the objective and/or constraints.

\noindent RTO and RBTO for realistic engineering applications confront
two challenges: (1) the theoretically infinite-dimensional design
vector; and (2) high-dimensional integration resulted from a large
number of random variables. Both lead to the curse of dimensionality,
which hinders or invalidates almost all RTO and RBTO methods. In RTO,
the objective or constraint functions are usually expressed by first
two moment properties, such as means and standard deviations, of certain
stochastic responses, describing the objective robustness or feasibility
robustness of a given topology. RBTO often contains probabilistic
constraint functions, which restrict the probability of failure regarding
certain failure mechanisms. Therefore, to solve a practical RTO or
RBTO problem using gradient-based algorithm, an efficient and accurate
method for statistical moments, reliability, and their sensitivity
analysis of random responses are in demands.

\noindent The fundamental problem rooted in statistical moment or
reliability analyses entails the evaluation of a high-dimensional
integral in the entire support of random inputs or its unknown subdomain,
respectively. In general, such an integral cannot be calculated analytically.
Direct numerical quadrature can be applied, but it is computationally
prohibitive when the number of random inputs exceeds three or four,
especially when the evaluation of a response function is carried out
by expensive finite element analyses (FEA). Existing approaches for
statistical moment and reliability analysis include the point estimate
method (PEM) \citep{huang2007}, Taylor series expansion or perturbation
method \citep{huang2007}, tensor product quadrature (TPQ) \citep{lee2009},
Neumann expansion method \citep{yamazaki88}, the first-order reliability
method or FORM-based methods \citep{kuschel1997,tu1999,du2004seq,chiralaksanakul2005,agarwal2006,liang2007},
polynomial chaos expansion (PCE) \citep{wang2006}, statistically
equivalent solution \citep{grigoriu91}, dimension-reduction method
\citep{xu04,xu05}, and others \citep{grigoriu02}. Their topology
sensitivities have relied mainly on two kinds of approaches: SIMP-based
approaches (solid isotropic material with penalization) \citep{amstutz2012topological}
and topology-derivative-based approaches \citep{novotny2003topological,allaire2005structural}.
The former is based on a fictitious density field representing a smooth
transition between material and empty, which requires regularization
procedures to get a clear topology. The latter introduces the topological
derivative concept which defines the derivative of functionals whose
variable is a geometrical domain with respect to singular topology
perturbation. The topological derivative concept is mathematically
rigorous and independent of the fictitious density field.

\noindent Nonetheless, three major concerns arise when evaluating
stochastic quantities and their sensitivity using existing approaches
or techniques. First, when applied to large-scale topology optimization
subject to a large number of random inputs, many of those methods
including Taylor series expansions, FORM-based methods, PEM, PCE,
TPQ, and dimension-reduction methods, etc. begin to be inapplicable
or inadequate. For example, although the Taylor series expansion,
FORM-based methods, and PEM are inexpensive and simple, they deteriorate
due to the lack of accuracy when the nonlinearity of a response function
is high and/or when the input uncertainty is large. PCE approximates
the random response via an infinite series of Hermite polynomials
of Gaussian variables (or others) and is popular in stochastic mechanics
in the last decades. Although truncated forms of PCE were extensively
used in practice \citep{zhang2017robust,kang2018reliability}, it
is easily succumbed to the curse of dimensionality due to astronomically
large numbers of terms or coefficients required to capture the interaction
effects between random inputs when applied to high-dimensional systems.
Rooted in the referential dimensional decomposition (RDD), the dimension-reduction
approximates a high dimensional function via a set of low dimensional
components, but it often results in sub-optimal estimations of the
original function, and thus its stochastic moments and the associated
reliability. Second, to evaluate the topology sensitivity of stochastic
quantities, many of the aforementioned methods may not be adequately
efficient and accurate. Most of those methods rely on a fictitious
density field, thus the sensitivities supplied are not the exact topology
sensitivity. Furthermore, many of them resort to repetitive stochastic
analyses especially for the sensitivity of reliability due to employed
finite-difference techniques, which restrain their computational efficiency.
Although Taylor series expansions, is able to perform stochastic sensitivities
analysis economically, its accuracy is usually deteriorated by inherited
errors from the associated stochastic analysis. Third, to the best
of the author's knowledge, in existing literature, there is no benchmark
example that provides analytical or semi-analytical solutions for
stochastic topology sensitivity analysis. A successful benchmark example
certainly calls for analytical expressions of stress, strain, or other
response functions in two domains - an original domain and a perforated
domain - subject to the same loads and supports. These analytical
expressions generally are not readily available even for simple domain
and ordinary load cases. Moreover, verifying the performance of a
certain method subject to high-dimensional random inputs often demands
the benchmark example carrying on complex loads to accommodate a large
number of random variables, which impede the implementation of analytical
solution of stochastic topology sensitivity. These difficulties result
in the lack of benchmark examples and make it impossible to verify
the accuracy of existing and new algorithms, especially for high-dimensional
cases.

\noindent This paper presents a novel framework for topology sensitivity
analysis of statistical moments and reliability for complex engineering
structures subject to a large number of random inputs. The framework,
designed for dealing high-dimensional random inputs, is grounded on
the polynomial dimensional decomposition (PDD), and thus it is capable
of approximating the high-dimensional stochastic responses in an efficient
and accurate manner. It also dovetails the deterministic topology
derivatives with PDD and provides stochastic sensitivities in the
exactly topological sense. For RTO, the proposed framework is endowed
with analytical expressions for topology sensitivities of the first
three stochastic moments. For RBTO, it supplies embedded Monte Carlo
Simulation (MCS) and a finite difference formulation to estimate topology
sensitivities of failure probability. Furthermore, the evaluation
of moments and/or reliability and their topology sensitivity is accomplished
concurrently from only a single stochastic analysis. It is noteworthy
that two benchmark examples, which provide analytical/semi-analytical
topology sensitivity of moments and reliability, are developed for
the calibration of stochastic topology sensitivity algorithms. The
first example contains only two random variables but provides analytical
expressions for moments, reliability, and their topology sensitivities.
The second one accommodates 53 random variables, whereas the analytical
expressions provided can be easily expanded to any positive number
of random variables. The rest of this paper is organized as follows.
Section 2 formally defines general RTO and RBTO problems, including
a concomitant mathematical statement. Section 3 starts with a brief
exposition of the polynomial dimensional decomposition and associated
approximations, which result in explicit formulae for the first two
moments and an embedded MCS formulation for the reliability of a generic
stochastic response. Section 4 revisits the definition of topology
derivative and describes the new framework of stochastic topology
sensitivity analysis, which integrates PDD and deterministic topological
derivative as well as numerical procedures for topology sensitivities
of both stochastic moment and reliability. The calculation of PDD
expansion coefficients is briefly described in Section 5. Section
6 presents three numerical examples. Two benchmark examples are developed
to probe the accuracy and computational efforts of the proposed method.
One three-dimensional bracket is used to demonstrate the feasibility
of the new method for practical engineering applications. Finally,
conclusions are drawn in Section 7.

\section{Stochastic topology design problems\label{sec2}}

In the presence of uncertainties, a topology optimization problem
can include robust, probabilistic, or non-probabilistic constraints.
For RTO, both objective and constraint functions may involve the first
two moment properties for the assessment of robustness \citep{du2000}.
Whereas for RBTO, probabilistic functions are often embedded as constraints
to restrict the failure probability and achieve a high confidence
level on design \citep{enevoldsen1994,tu1999}. Nonetheless, the typical
RTO and RBTO problems interested in this paper are often formulated
as the following mathematical programming problems

\textcolor{black}{
\begin{eqnarray}
\min_{\Omega\subseteq D} &  & c_{0}(\Omega):=w_{1}\frac{{\displaystyle \mathbb{E}}\left[y_{0}(\Omega,\mathbf{X})\right]}{\mu_{0}^{*}}+w_{2}\frac{\sqrt{\mathrm{var}\left[y_{0}(\Omega,\mathbf{X})\right]}}{\sigma_{0}^{*}},\nonumber \\
\mbox{subject to} &  & c_{k}(\Omega):=\alpha_{k}\sqrt{\mathrm{\mathrm{var}}\left[y_{k}(\Omega,\mathbf{X})\right]}-{\displaystyle \mathbb{E}}\left[y_{k}(\Omega,\mathbf{X})\right]\le0;\ \ k=1,\cdots,K\label{eq:rto}
\end{eqnarray}
}

and

\textcolor{black}{
\begin{eqnarray}
\min_{\Omega\subseteq D} &  & c_{0}(\Omega):=w_{1}\frac{{\displaystyle \mathbb{E}}\left[y_{0}(\Omega,\mathbf{X})\right]}{\mu_{0}^{*}}+w_{2}\frac{\sqrt{\mathrm{var}\left[y_{0}(\Omega,\mathbf{X})\right]}}{\sigma_{0}^{*}},\nonumber \\
\mbox{subject to} &  & c_{k}(\Omega):=P\left[\mathbf{X}\in\Omega_{F,k}\right]\le p_{k};\ \ k=1,\cdots,K,\label{eq:rbto}
\end{eqnarray}
}respectively, where $D\subset\mathbb{R}^{3}$ is a bounded domain
in which all admissible topology design $\Omega$ are included; $\mathbf{X}:=(X_{1},\cdots,X_{N})^{T}\in\mathbb{R}^{N}$
is an $N$-dimensional random input vector completely defined by a
family of joint probability density functions $\{f_{\mathbf{X}}(\mathbf{x}),\:\mathbf{x}\in\mathbb{R}^{N}\}$
on the probability triple $(\Omega_{\mathbf{X}},\mathcal{F},P)$,
where $\Omega_{\mathbf{X}}$ is the sample space; $\mathcal{F}$ is
the $\sigma$-field on $\Omega_{\mathbf{X}}$; $P$ is the probability
measure associated with probability density $f_{\mathbf{X}}(\mathbf{x})$;
$\Omega_{F,k}$ is the $k$th failure domain defined by response function
$y_{k}(\Omega,\mathbf{X})$; $0<p_{k}<1$ expresses target failure
probabilities; $w_{1}\in\mathbb{R}_{0}^{+}$ and $w_{2}\in\mathbb{R}_{0}^{+}$
are two non-negative, real-valued weights, satisfying $w_{1}+w_{2}=1$,
$\mu_{0}^{*}\in\mathbb{R}\setminus\{0\}$ and $\sigma_{0}^{*}\in\mathbb{R}^{+}$
are two non-zero, real-valued scaling factors; $\alpha_{k}\in\mathbb{R}^{+}$,
$k=0,1,\cdots,K$, are positive, real-valued constants associated
with the probabilities of constraint satisfaction; $\mathbb{E}$ and
$\mathrm{var}$ are expectation operator and variance operator, respectively,
with respect to the probability measure $P$. The evaluation of both
$\mathbb{E}$ and $\mathrm{var}$ on certain random response demands
statistical moment analysis \citep{rosenblueth1981,hong1998,Kleiber92,rahman01p,evans1967,yamazaki88,grigoriu91,xu04,xu05,grigoriu02},
which is not unduly difficult. By contrast, the evaluation of probabilistic
constraint functions in RBTO, generally more complicated than $\mathbb{E}$
and $\mathrm{var}$, is obtained from

\begin{equation}
c_{k}(\Omega):=P\left[\mathbf{X}\in\Omega_{F,k}\right]=\int_{\Omega_{F,k}}f_{\mathbf{X}}(\mathbf{x})d\mathbf{x}=\int_{\mathbb{R}^{N}}I_{\Omega_{F,k}}(\Omega,\mathbf{x})d\mathbf{x}:=\mathbb{E}\left[I_{\Omega_{F,k}}(\Omega,\mathbf{X})\right]
\end{equation}
which represents a failure probability from \textit{reliability analysis}
\citep{cornell1969,madsen1986,hasofer1974,fiessler1979,hohenbichler1981,breitung1984,der1987,wu1990,wu1987,liu1991}.
The indicator function $I_{\Omega_{F,k}}(\Omega,\mathbf{x})=1$ when
\textbf{$\mathbf{x}\in\Omega_{F,k}$ }and \textit{zero} otherwise.
For component-level RBTO, the failure domain, often adequately described
by a single performance function $y_{k}(\Omega,\mathbf{x})$, and
component reliability analysis are relatively simple. Whereas, interdependent
performance functions $y_{k}^{(q)}(\Omega,\mathbf{x}),\:q=1,2,\cdots$,
are required for a system-level (series, parallel, or general) RBTO,
leading to a highly complex failure domain and huge computational
demand for system reliability analysis.

\section{Polynomial dimensional decomposition method and uncertainty quantification
\label{sec3}}

\subsection{Polynomial dimensional decomposition}

\noindent Consider a multivariate stochastic response $y(\Omega,\mathbf{X})$
of certain topology design $\Omega$ subject to random input vector\textcolor{black}{{}
$\mathbf{X}=\{X_{1},\text{\ensuremath{\cdots}},X_{N}\}^{T}$}, representing
any of the performance function $y_{k}$ in Eq. \eqref{eq:rto} or
\eqref{eq:rbto}\textcolor{black}{.}\textcolor{red}{{} }\textcolor{black}{Let
$\mathcal{L}_{2}(\Omega_{\mathbf{X}},\mathcal{F},P)$ be} a Hilbert
space of square-integrable functions $y$ with a probability mea\textcolor{black}{sure
$f_{\mathbf{X}}(\mathbf{x})d\mathbf{x}$ sup}ported on $\mathbb{R}^{N}$.
Assuming independent components of $\mathbf{X}$, the PDD expansion
of function $y$ generates a hierarchical representation\citep{rahman2008,rahman2009}

\begin{equation}
y(\Omega,\mathbf{X})=y_{\emptyset}(\Omega)+\sum_{\emptyset\ne u\subseteq\{1,\cdots,N\}}\sum_{\mathbf{j}_{|u|}\in\mathbb{N}^{|u|}}C_{u\mathbf{j}_{|u|}}(\Omega)\psi_{u\mathbf{j}_{|u|}}(\mathbf{X}_{u};\Omega),\label{eq:PDD}
\end{equation}
of the original performance function, in terms of an infinite number
of multivariate orthonormal basis \citep{rahman2008,rahman2009} $\psi_{u\mathbf{j}_{|u|}}(\mathbf{X}_{u};\Omega):=\prod_{p=1}^{|u|}\psi_{i_{p}j_{p}}(X_{i};\Omega)$
in \textcolor{black}{$\mathcal{L}_{2}(\Omega_{\mathbf{X}},\mathcal{F},P)$},
where $\mathbf{j}_{|u|}=(j_{1},\cdots,j_{|u|})\in\mathbb{N}^{|u|}$
is a $|u|$-dimensional multi-index; $y_{\phi}(\Omega)$ contributes
the constant component; for $|u|=1$, $C_{u\mathbf{j}_{|u|}}(\Omega)\psi_{u\mathbf{j}_{|u|}}(\mathbf{X}_{u};\Omega)$
commits all univariate component functions representing the individual
contribution to $y(\Omega,\mathbf{X})$ from each single input variable;
for $|u|=2$, it brings in all bivariate component functions embodying
the cooperative influence of any two input variables; and for $|u|=S$,
it admits $S$-variate component functions quantifying the interactive
effects of any $S$ input variables. For most performance functions
in engineering applications, a truncated version of Eq. \eqref{eq:PDD}
is often accurate enough by retaining, at most, the interactive effects
of $S<N$ input variables and $m$th order polynomials, 
\begin{align}
\tilde{y}_{S,m}(\Omega,\mathbf{X}) & =y_{\emptyset}(\Omega)+{\displaystyle \sum_{{\textstyle {\emptyset\ne u\subseteq\{1,\cdots,N\}\atop 1\le|u|\le S}}}}\sum_{{\textstyle {\mathbf{j}_{|u|}\in\mathbb{N}^{|u|}\atop \left\Vert \mathbf{j}_{|u|}\right\Vert _{\infty}\le m}}}\!\!\!\!\!\!C_{u\mathbf{j}_{|u|}}(\Omega)\psi_{u\mathbf{j}_{|u|}}(\mathbf{X}_{u};\Omega),\label{eq:T-PDD}
\end{align}
where 
\begin{equation}
y_{\emptyset}(\Omega)=\int_{\mathbb{R}^{N}}y(\mathbf{x},\Omega)f_{\mathbf{X}}(\mathbf{x})d\mathbf{x}\label{eq:y0}
\end{equation}
and 
\begin{eqnarray}
C_{u\mathbf{j}_{|u|}}(\Omega): & = & \int_{\mathbb{R}^{N}}y(\mathbf{x},\Omega)\psi_{u\mathbf{j}_{|u|}}(\mathbf{\mathbf{x}}_{u};\Omega)f_{\mathbf{X}}(\mathbf{x})d\mathbf{x},\ \ \ \;\emptyset\ne u\subseteq\{1,\cdots,N\},\;\mathbf{j}_{|u|}\in\mathbb{N}^{|u|},\label{eq:cuju}
\end{eqnarray}
are referred to as expansion coefficients of PDD expansion \eqref{eq:PDD}
or truncated PDD approximation \eqref{eq:T-PDD}. The untruncated
PDD expansion in Eq. \eqref{eq:PDD} employs an orthogonal polynomial
basis and exactly represents the response function, it can be easily
refer that it is equivalent to PCE when the basis used is same. However,
the PDD expansion provides a hierarchical representation by classifying
the interaction between random inputs, which is a key to alleviate
the course of dimensionality when applying its truncated version.
For $S>0$ and $m>0$, Eq. \eqref{eq:T-PDD} retains interactive effects
among at most $S$ input variables $X_{i_{1}},\cdots,X_{i_{S}}$,
$1\leq i_{1}<\cdots<i_{S}\leq N$ and $m$th order polynomial nonlinearity
in $y$, thus leading to the so-called $S$-variate, $m$th-order
PDD approximation. When $S\to N$ and $m\to\infty$, $\tilde{y}_{S,m}$
converges to $y$ in the mean-square sense and engenders a sequence
of hierarchical and convergent approximations of $y$. Based on the
dimensional structure and nonlinearity of a stochastic response, the
truncation parameters $S$ and $m$ can be chosen correspondingly.
The higher the values of $S$ and $m$ permit the higher the accuracy,
but also endow the computational cost of an $S$th-order polynomial
computational complexity \citep{rahman2008,rahman2009}. Henceforth,
the $S$-variate, $m$th-order PDD approximation will be simply referred
to as \emph{truncated PDD approximation} in this paper.

\subsection{Stochastic moment analysis}

\noindent For an arbitrary random response of certain topology design
$\Omega$, let $m^{(r)}(\Omega):=\mathbb{E}[y^{r}(\Omega,\mathbf{X})]$,
if it exists, denote the raw moment of $y$ of order $r$, where $r\in\mathbb{N}$.
Let $\tilde{m}^{(r)}(\Omega):=\mathbb{E}[\tilde{y}_{S,m}^{r}(\Omega,\mathbf{X})]$
denote the raw moment of $\tilde{y}_{S,m}$ of order $r$, given an
$S$-variate, $m$th-order PDD approximation $\tilde{y}_{S,m}(\Omega,\mathbf{X})$
of $y(\Omega,\mathbf{X})$. The analytical expressions or explicit
formulae for estimating the moments using PDD approximations are described
as follows. Applying the expectation operator on $\tilde{y}_{S,m}(\Omega,\mathbf{X})$
and $\tilde{y}_{S,m}^{2}(\Omega,\mathbf{X})$, the first moment or
mean \citep{rahman10} 
\begin{equation}
\tilde{m}_{S,m}^{(1)}(\Omega):=\mathbb{E}\left[\tilde{y}_{S,m}(\Omega,\mathbf{X})\right]=y_{\emptyset}(\Omega)=\mathbb{E}\left[y(\Omega,\mathbf{X})\right]=:m^{(1)}(\Omega)\label{mom1}
\end{equation}
of the $S$-variate, $m$th-order PDD approximation is simply the
constant component in Eq. \eqref{eq:T-PDD}, whereas the second moment
\citep{rahman10} 
\begin{equation}
\tilde{m}_{S,m}^{(2)}(\Omega):=\mathbb{E}\left[\tilde{y}_{S,m}^{2}(\Omega,\mathbf{X})\right]=y_{\emptyset}^{2}(\Omega)+{\displaystyle \sum_{{\textstyle {\emptyset\ne u\subseteq\{1,\cdots,N\}\atop 1\le|u|\le S}}}}\sum_{{\textstyle {\mathbf{j}_{|u|}\in\mathbb{N}^{|u|}\atop \left\Vert \mathbf{j}_{|u|}\right\Vert _{\infty}\le m}}}C_{u\mathbf{j}_{|u|}}^{2}(\Omega)\label{mom2}
\end{equation}
is expressed as the sum of squares of all expansion coefficients of
$\tilde{y}_{S,m}(\Omega,\mathbf{X})$. It is straightforward that
the estimation of the second moment evaluated by Eq. \eqref{mom2}
approaches the exact second moment 
\begin{equation}
m^{(2)}(\Omega):=\mathbb{E}\left[y^{2}(\Omega,\mathbf{X})\right]=y_{\emptyset}^{2}(\Omega)+{\displaystyle \sum_{\emptyset\ne u\subseteq\{1,\cdots,N\}}}\sum_{{\textstyle {\mathbf{j}_{|u|}\in\mathbb{N}^{|u|}\atop }}}C_{u\mathbf{j}_{|u|}}^{2}(\Omega)
\end{equation}
of $y$ when $S\to N$ and $m\to\infty$. The mean-square convergence
of $\tilde{y}_{S,m}$ is ensured as its component functions will contain
all required bases of the corresponding Hilbert spaces. Furthermore,
the variance of $\tilde{y}_{S,m}(\Omega,\mathbf{X})$ is also mean-square
convergent.\\

\subsection{Reliability analysis}

The RBTO problem defined in Eq. \eqref{eq:rbto} requires not only
stochastic moment analysis but also evaluations of the probabilistic
constraints, that is, the failure probability 
\begin{equation}
P_{F}=P\left[\mathbf{X}\in\Omega_{F,k}\right]=\int_{\Omega_{F,k}}f_{\mathbf{X}}(\mathbf{x})d\mathbf{x}=\int_{\mathbb{R}^{N}}I_{\Omega_{F,k}}(\Omega,\mathbf{x})d\mathbf{x}:=\mathbb{E}\left[I_{\Omega_{F,k}}(\Omega,\mathbf{X})\right]
\end{equation}
of a certain topology design $\Omega$ with respect to certain failure
set $\Omega_{F,k}$. In which, the indicator function $I_{\Omega_{F,k}}(\Omega,\mathbf{x})=1$
when \textbf{$\mathbf{x}\in\Omega_{F,k}$ }and \textit{zero} otherwise.
For component-level RBTO, the failure set is often adequately characterized
by a single performance function $y_{k}(\Omega,\mathbf{x})$ as $\Omega_{F,k}:=\{\mathbf{x}:y_{k}(\Omega,\mathbf{x})<0\}$.
Whereas for a system-level RBTO, it is usually described by multiple,
interdependent performance functions $y_{k}^{(q)}(\Omega,\mathbf{x}),\:q=1,2,\cdots$,
leading, for example, to $\Omega_{F,k}:=\{\mathbf{x}:\cup_{q}y_{k}^{(q)}(\Omega,\mathbf{x}),<0\}$
and $\Omega_{F,k}:=\{\mathbf{x}:\cap_{q}y_{k}^{(q)}(\Omega,\mathbf{x}),<0\}$
for series and parallel systems, respectively.\textcolor{black}{{}
Let $\tilde{\Omega}_{F,k}:=\{\mathbf{x}:\tilde{y}_{S,m}(\mathbf{x})<0\}$
or $\tilde{\Omega}_{F,k}:=\{\mathbf{x}:\cup_{q}\tilde{y}_{S,m}^{(q)}(\mathbf{x})<0\}$
or $\tilde{\Omega}_{F,k}:=\{\mathbf{x}:\cap_{q}\tilde{y}_{S,m}^{(q)}(\mathbf{x})<0\}$
be an approximate failure set as a result of $S$-variate, $m$th-order
PDD approximations $\tilde{y}_{S,m}(\mathbf{X})$ of $y(\mathbf{X})$
or $\tilde{y}_{S,m}^{(q)}(\mathbf{X})$ of $y^{(q)}(\mathbf{X})$.
Then the embedded MCS estimate of the failure probability $P_{F}$
is }

\begin{equation}
\tilde{P}_{F}=\mathbb{E}_{\mathbf{d}}\left[I_{\tilde{\Omega}_{F,k}}(\Omega,\mathbf{X})\right]=\lim\limits _{L\rightarrow\infty}\frac{1}{L}\sum\limits _{l=1}^{L}I_{\tilde{\Omega}_{F,k}}(\Omega,\mathbf{x}^{(l)}),\label{Eq:pfmcs}
\end{equation}
where $L$ is the sample size, $\mathbf{x}^{(l)}$ is the $l$th realization
of $\mathbf{X}$, and $I_{\tilde{\Omega}_{F,k}}(\Omega,\mathbf{x})$,
equal to \emph{one} when $\mathbf{x}\in\tilde{\Omega}_{F,k}$ and
\emph{zero} otherwise, is an approximation of the indicator function
$I_{\Omega_{F,k}}(\Omega,\mathbf{x})$.

Note that the stochastic moment analysis and reliability analysis
for RTO and RBTO are quite similar to the ones in a general robust
design optimization (RDO) and reliability-based design optimization
(RBDO) \citep{ren2013robust,rahman2014novel,ren2015reliability} except
that the former is affiliated with certain topology designs $\Omega$.
However, topology sensitivity analysis of moments and reliability
is distinct from sensitivity analysis in RDO and RBDO due to the disparate
topology change associated, and is elaborated in the next section.

\section{Stochastic topology sensitivity analysis\label{sec4}}

To evaluate the topology sensitivity of a stochastic response, a new
framework is proposed here which dovetails PDD and deterministic topological
derivative. It relies fundamentally on the topology derivative \citep{td5,td6,td4,td1,td2,td3,L4,L6,L7}
of a deterministic objective function $y(\Omega)$. The new method
provides closed-form solutions and an embedded MCS formulation for
the topological derivative of stochastic moments and reliability,
respectively. Before presenting the new framework itself, a brief
revisit on the idea of topological derivative appears to be necessary
and should be convenient to those not yet familiar with the concept.

\subsection{Topological derivative - revisit}

\begin{figure}
\centering{}\includegraphics[scale=0.3]{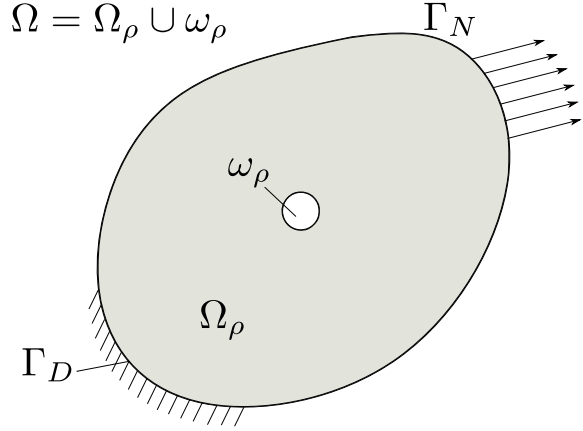}\caption{A perforated domain}
\label{Fig:domain-omega_rho} 
\end{figure}

Pioneered by Schumacher\citep{shumacher1995topologieoptimierung},
Sokolowski and Zochowski \citep{sokolowski1999topological,sokolowski2009topological},
and Garreau et al. \citep{garreau2001topological}, the \textit{topological
derivative} measures the change of a performance functional when an
infinitesimal hole is introduced in the reference domain in which
a boundary-value problem is defined. For a given reference domain
$\Omega\subset\mathbb{R}^{n}$, a point $\mathbf{\boldsymbol{\xi}}_{0}\in\Omega$,
and a hole $\omega\in\mathbb{R}^{n}$ with the radius of $1$, a translated
and rescaled hole can be defined by $\omega_{\rho}=\boldsymbol{\xi}_{0}+\rho\omega,\ \forall\rho>0$
and the perforated domain is $\Omega_{\rho}=\Omega\backslash\bar{\omega}_{\rho}$
as shown in Fig. \ref{Fig:domain-omega_rho}. 

For a small $\rho>0$, if $y(\Omega_{\rho})$ admits the topological
asymptotic expansion 
\begin{equation}
y(\Omega_{\rho})=y(\Omega)+\rho^{n}D_{T}y(\boldsymbol{\xi}_{0})+o(\rho^{n}),\label{eq:asymptotic}
\end{equation}
then $D_{T}y(\boldsymbol{\xi}_{0})$ is called the topological derivative
at point $\boldsymbol{\xi}_{0}$ and is applicable to general boundary
value problems including the linear elastic system

\begin{equation}
\begin{cases}
\bm{\nabla}\cdot(\mathbb{C}:\bm{\nabla}\bm{u})=\bm{0} & \textrm{in }\Omega\\
\mathbf{u}=\mathbf{\bar{u}} & \textrm{on }\Gamma_{D}\\
\mathbf{n}\cdot(\mathbb{C}:\bm{\nabla}\bm{u})=:\boldsymbol{t}=\bar{\boldsymbol{t}} & \textrm{on }\Gamma_{N}
\end{cases}.\label{eq:BVP}
\end{equation}
where $\mathbb{C}$ is the elastic tensor, $\Gamma_{D}$ and $\Gamma_{N}$
denote Dirichlet boundary and Neumann boundary of $\Omega$, respectively.
The topological asymptotic expansion \eqref{eq:asymptotic} contains
two performance functions $y(\Omega)$ and $y(\Omega_{\rho})$. The
former is related to the reference domain $\Omega$ and evaluated
by solving \eqref{eq:BVP}, whereas the latter is affiliated with
the perforated domain $\Omega_{\rho}$ and the associated boundary
value problem

\begin{equation}
\begin{cases}
\bm{\nabla}\cdot\left[\mathbb{C}:\bm{\nabla}\left(\bm{u}+\hat{\boldsymbol{u}}\right)\right]=\bm{0} & \textrm{in }\Omega_{\rho}\\
\bm{u}+\hat{\boldsymbol{u}}=\bar{\bm{u}} & \textrm{on }\Gamma_{D}\\
\bm{n}\cdot\left[\mathbb{C}:\bm{\nabla}\left(\bm{u}+\hat{\boldsymbol{u}}\right)\right]=:\boldsymbol{t}+\hat{\boldsymbol{t}}=\bar{\bm{t}} & \textrm{on }\Gamma_{N}\\
\boldsymbol{t}+\hat{\boldsymbol{t}}=\bm{0} & \textrm{on }-\partial\omega_{\rho}
\end{cases}\label{eq:BVP-perforated}
\end{equation}
where the Neumann type condition is prescribed on $-\partial\omega_{\rho}$,
i.e., the boundary $\partial\omega_{\rho}$ with the opposite normal
vector. Comparing Eq. \eqref{eq:BVP} and Eq. \eqref{eq:BVP-perforated},
it concludes that $\hat{\boldsymbol{u}}=\boldsymbol{0}$ on $\Gamma_{D}$
and $\hat{\boldsymbol{t}}=\boldsymbol{0}$ on $\Gamma_{N}$. Moreover,
it was proved that $\hat{\boldsymbol{u}}+o\left(\rho\right)$, where
$o\left(\rho\right)$ is the reminder of higher order compared to
$\rho$, is the solution of the following \textit{external problem}
\citep{garreau2001topological} 
\begin{equation}
\begin{cases}
\bm{\nabla}\cdot(\mathbb{C}:\bm{\nabla}\hat{\boldsymbol{u}})=\bm{0} & \textrm{in }\mathbb{R}^{n}\backslash\overline{\omega_{\rho}}\\
\bm{n}\mathbf{\cdot}\left(\mathbb{C}:\bm{\nabla}\hat{\boldsymbol{u}}\right)=:\hat{\boldsymbol{t}}=\bm{n}\cdot\mathbb{C}:\bm{\nabla}\boldsymbol{u}\left(\bm{\xi}_{0}\right) & \textrm{on }-\partial\omega_{\rho}
\end{cases},\label{eq:BVP-external}
\end{equation}
as $\rho\rightarrow0$. Solutions $\hat{\boldsymbol{u}}$ for various
cases of isotropic elasticity are summarized in Table \ref{Table:1},
for more details and an easy solution utilizing the \textit{Eshelby
tensor}, refer to \ref{appendix1}.

Both $y(\Omega)$ and $y(\Omega_{\rho})$ admit a general class of
performance functions. Consider the compliance of the structure as
the performance functional, $y(\Omega):=\int_{\Gamma_{D}\cup\Gamma_{N}}\boldsymbol{u}\cdot\boldsymbol{t}{\rm d}\Gamma$,
which can be augmented by a Lagrange multiplier $\boldsymbol{\lambda}$
to introduce the governing equation as follows, 
\begin{equation}
y(\Omega):=\int_{\Gamma_{D}\cup\Gamma_{N}}\boldsymbol{u}\cdot\boldsymbol{t}{\rm d}\Gamma=\int_{\Gamma_{D}\cup\Gamma_{N}}\boldsymbol{u}\cdot\boldsymbol{t}{\rm d}\Gamma+\int_{\Omega}\boldsymbol{\lambda}\cdot\left[\bm{\nabla}\cdot(\mathbb{C}:\bm{\nabla}\bm{u})\right]{\rm d}\Omega,
\end{equation}
by noticing $\boldsymbol{u}$ being the solution of Eq. \eqref{eq:BVP}
in advance, where $\boldsymbol{\lambda}$ can be any kinematically
admissible field that meets appropriate smoothness requirements. Similarly
for the perforated domain, 
\begin{equation}
y(\Omega_{\rho}):=\int_{\Gamma_{D}\cup\Gamma_{N}}\left(\bm{u}+\hat{\boldsymbol{u}}\right)\cdot\left(\boldsymbol{t}+\hat{\boldsymbol{t}}\right){\rm d}\Gamma+\int_{\Omega_{\rho}}\boldsymbol{\lambda}\cdot\left[\bm{\nabla}\cdot(\mathbb{C}:\bm{\nabla}\left(\bm{u}+\hat{\boldsymbol{u}}\right))\right]{\rm d}\Omega.
\end{equation}

The change of compliance after perforation

\begin{align}
y(\Omega_{\rho})-y(\Omega)= & \int_{\Gamma_{D}\cup\Gamma_{N}}\left(\bm{u}\cdot\hat{\boldsymbol{t}}+\hat{\boldsymbol{u}}\cdot\boldsymbol{t}+\hat{\boldsymbol{u}}\cdot\hat{\boldsymbol{t}}\right){\rm d}\Gamma+\int_{\Omega_{\rho}}\boldsymbol{\lambda}\cdot\left[\bm{\nabla}\cdot(\mathbb{C}:\bm{\nabla}\hat{\boldsymbol{u}})\right]{\rm d}\Omega-\int_{\omega_{\rho}}\boldsymbol{\lambda}\cdot\left[\bm{\nabla}\cdot(\mathbb{C}:\bm{\nabla}\boldsymbol{u})\right]{\rm d}\Omega\nonumber \\
= & \int_{\Gamma_{D}\cup\Gamma_{N}}\left(\bm{u}\cdot\hat{\boldsymbol{t}}+\hat{\boldsymbol{u}}\cdot\boldsymbol{t}\right){\rm d}\Gamma+\int_{\Omega_{\rho}}\boldsymbol{\lambda}\cdot\left[\bm{\nabla}\cdot(\mathbb{C}:\bm{\nabla}\hat{\boldsymbol{u}})\right]{\rm d}\Omega-\int_{\omega_{\rho}}\boldsymbol{\lambda}\cdot\left[\bm{\nabla}\cdot(\mathbb{C}:\bm{\nabla}\boldsymbol{u})\right]{\rm d}\Omega,
\end{align}
employing $\hat{\boldsymbol{u}}\rightarrow\boldsymbol{0}$ on $\Gamma_{D}$
and $\hat{\boldsymbol{t}}\rightarrow\boldsymbol{0}$ on $\Gamma_{N}$
as $\rho\rightarrow0$. Integrate the second term of the above equation
by parts twice and the third term one time, meanwhile applying divergence
theorem, 
\begin{align}
y(\Omega_{\rho})-y(\Omega)= & \int_{\Gamma_{D}\cup\Gamma_{N}}\left(\bm{u}\cdot\hat{\boldsymbol{t}}+\hat{\boldsymbol{u}}\cdot\boldsymbol{t}\right){\rm d}\Gamma+\int_{\Gamma_{D}\cup\Gamma_{N}\cup-\partial\omega_{\rho}}\boldsymbol{\lambda}\cdot\hat{\boldsymbol{t}}{\rm d}\Gamma-\int_{\Omega_{\rho}}\boldsymbol{\nabla}\boldsymbol{\lambda}:\mathbb{C}:\bm{\nabla}\hat{\boldsymbol{u}}{\rm d}\Omega-\int_{\partial\omega_{\rho}}\boldsymbol{\lambda}\cdot\boldsymbol{t}{\rm d}\Gamma+\int_{\omega_{\rho}}\boldsymbol{\nabla}\boldsymbol{\lambda}:\mathbb{C}:\bm{\nabla}\boldsymbol{u}{\rm d}\Omega\nonumber \\
= & \int_{\Gamma_{D}\cup\Gamma_{N}}\left(\bm{u}\cdot\hat{\boldsymbol{t}}+\hat{\boldsymbol{u}}\cdot\boldsymbol{t}+\boldsymbol{\lambda}\cdot\hat{\boldsymbol{t}}\right){\rm d}\Gamma-\int_{\partial\omega_{\rho}}\boldsymbol{\lambda}\cdot\hat{\boldsymbol{t}}{\rm d}\Gamma-\int_{\Gamma_{D}\cup\Gamma_{N}\cup-\partial\omega_{\rho}}\hat{\boldsymbol{u}}\cdot\left(\boldsymbol{n}\cdot\mathbb{C}:\bm{\nabla}\boldsymbol{\lambda}\right){\rm d}\Gamma+\int_{\Omega_{\rho}}\hat{\boldsymbol{u}}\cdot\left[\bm{\nabla}\cdot\left(\mathbb{C}:\boldsymbol{\nabla}\boldsymbol{\lambda}\right)\right]{\rm d}\Omega\nonumber \\
 & -\int_{\partial\omega_{\rho}}\boldsymbol{\lambda}\cdot\boldsymbol{t}{\rm d}\Gamma+\int_{\omega_{\rho}}\boldsymbol{\nabla}\boldsymbol{\lambda}:\mathbb{C}:\bm{\nabla}\boldsymbol{u}{\rm d}\Omega\nonumber \\
= & \int_{\Gamma_{D}}\left(\bm{u}+\boldsymbol{\lambda}\right)\cdot\hat{\boldsymbol{t}}{\rm d}\Gamma+\int_{\Gamma_{N}}\hat{\boldsymbol{u}}\cdot\boldsymbol{t}{\rm d}\Gamma-\int_{\partial\omega_{\rho}}\boldsymbol{\lambda}\cdot\left(\boldsymbol{t}+\hat{\boldsymbol{t}}\right){\rm d}\Gamma-\int_{\Gamma_{N}}\hat{\boldsymbol{u}}\cdot\left(\boldsymbol{n}\cdot\mathbb{C}:\bm{\nabla}\boldsymbol{\lambda}\right){\rm d}\Gamma+\int_{\Omega_{\rho}}\hat{\boldsymbol{u}}\cdot\left[\bm{\nabla}\cdot\left(\mathbb{C}:\boldsymbol{\nabla}\boldsymbol{\lambda}\right)\right]{\rm d}\Omega\nonumber \\
 & +\int_{\partial\omega_{\rho}}\hat{\boldsymbol{u}}\cdot\left(\boldsymbol{n}\cdot\mathbb{C}:\bm{\nabla}\boldsymbol{\lambda}\right){\rm d}\Gamma+\int_{\omega_{\rho}}\boldsymbol{\nabla}\boldsymbol{\lambda}:\mathbb{C}:\bm{\nabla}\boldsymbol{u}{\rm d}\Omega\nonumber \\
= & \int_{\Gamma_{D}}\left(\bm{u}+\boldsymbol{\lambda}\right)\cdot\hat{\boldsymbol{t}}{\rm d}\Gamma+\int_{\Gamma_{N}}\hat{\boldsymbol{u}}\cdot\left(\boldsymbol{t}-\boldsymbol{n}\cdot\mathbb{C}:\bm{\nabla}\boldsymbol{\lambda}\right){\rm d}\Gamma+\int_{\Omega_{\rho}}\hat{\boldsymbol{u}}\cdot\left[\bm{\nabla}\cdot\left(\mathbb{C}:\boldsymbol{\nabla}\boldsymbol{\lambda}\right)\right]{\rm d}\Omega+\int_{\partial\omega_{\rho}}\hat{\boldsymbol{u}}\cdot\left(\boldsymbol{n}\cdot\mathbb{C}:\bm{\nabla}\boldsymbol{\lambda}\right){\rm d}\Gamma\nonumber \\
 & +\int_{\omega_{\rho}}\boldsymbol{\nabla}\boldsymbol{\lambda}:\mathbb{C}:\bm{\nabla}\boldsymbol{u}{\rm d}\Omega,
\end{align}
noticing $\hat{\boldsymbol{u}}=\boldsymbol{0}$ on $\Gamma_{D}$,
$\hat{\boldsymbol{t}}=\boldsymbol{0}$ on $\Gamma_{N}$, $\boldsymbol{t}+\hat{\boldsymbol{t}}=\boldsymbol{0}$
on $\partial\omega_{\rho}$, and $\boldsymbol{n}$ is always the normal
of the current integration surface during the above derivation. Take
$\boldsymbol{\lambda}$ as the displacement solution of the following
adjoint problem

\begin{equation}
\begin{cases}
\bm{\nabla}\cdot(\mathbb{C}:\boldsymbol{\nabla}\boldsymbol{\lambda})=0 & \textrm{in }\Omega\\
\boldsymbol{\lambda}=-\bar{\bm{u}} & \textrm{on }\Gamma_{D}\\
\bm{n}\mathbf{\cdot(\mathbb{C}:\boldsymbol{\nabla}\boldsymbol{\lambda})}=\bar{\bm{t}} & \textrm{on }\Gamma_{N}
\end{cases}\label{eq:BVP-top-adjoint-1}
\end{equation}
and apply the solution $\hat{\boldsymbol{u}}$ on $\partial\omega_{\rho}$
of the external problem for the three-dimensional case in Table \ref{Table:1},
we have

\begin{align}
y(\Omega_{\rho})-y(\Omega)= & \int_{\omega_{\rho}}\boldsymbol{\nabla}\boldsymbol{\lambda}:\mathbb{C}:\bm{\nabla}\boldsymbol{u}{\rm d}\Omega+\int_{\partial\omega_{\rho}}\hat{\boldsymbol{u}}\cdot\left(\boldsymbol{n}\cdot\mathbb{C}:\bm{\nabla}\boldsymbol{\lambda}\right){\rm d}\Gamma\nonumber \\
= & \frac{4\pi\rho^{3}}{3}\left(\mathbb{C}^{-1}:\tilde{\bm{\sigma}}\right):\bm{\sigma}+\rho\int_{\partial\omega_{\rho}}\left(\frac{a-b}{3}\mathrm{tr}\left(\boldsymbol{\sigma}\right)\boldsymbol{n}+b\boldsymbol{n}\cdot\boldsymbol{\sigma}\right)\cdot\left(\boldsymbol{n}\cdot\tilde{\bm{\sigma}}\right){\rm d}\Gamma\nonumber \\
= & \frac{4\pi\rho^{3}}{3}\tilde{\bm{\sigma}}:\mathbb{C}^{-1}:\bm{\sigma}+\rho\left[b\left(\bm{\tilde{\sigma}}\bm{\cdot\sigma}\right):\int_{\partial\omega_{\rho}}\bm{n}\bm{n}{\rm d}\Gamma+\frac{a-b}{3}{\rm tr}\left(\bm{\sigma}\right)\bm{\tilde{\sigma}:}\int_{\partial\omega_{\rho}}\bm{n}\bm{n}{\rm d}\Gamma\right]\nonumber \\
= & \frac{4\pi\rho^{3}}{3}\left[\tilde{\bm{\sigma}}:\mathbb{C}^{-1}:\bm{\sigma}+\left[b\bm{\tilde{\sigma}}:\mathbb{I}:\bm{\sigma}+\frac{a-b}{3}\bm{\tilde{\sigma}:}\bm{\delta}\bm{\delta}:\bm{\sigma}\right]\right],
\end{align}
identifying $\int_{\partial\omega_{\rho}}\bm{n}\bm{n}=\frac{4\pi\rho^{2}}{3}\bm{\delta}$
for the three-dimensional case, where $\bm{\delta}$ is the second-order
unit tensor, $\mathbb{I}$ is the fourth-order identity tensor, and
$\tilde{\bm{\sigma}}=\mathbb{C}:\boldsymbol{\nabla}\boldsymbol{\lambda}$
is the stress solution at $\boldsymbol{\xi}_{0}$ of the adjoint problem.
Further calculations lead to 
\begin{align}
y(\Omega_{\rho})-y(\Omega)= & \frac{4\pi\rho^{3}}{3}\tilde{\bm{\sigma}}:\left[\left(\frac{(1+\nu)}{E}+b\right)\mathbb{I}+\left(\frac{a-b}{3}-\frac{\nu}{E}\right)\bm{\delta}\bm{\delta}\right]:\bm{\sigma}\nonumber \\
= & 4\pi\rho^{3}\frac{1-\nu}{2E(7-5\nu)}\tilde{\bm{\sigma}}:\left[10(1+\nu)\mathbb{I}-\left(5\nu+1\right)\bm{\delta}\bm{\delta}\right]:\bm{\sigma}\nonumber \\
:= & \rho^{3}\tilde{\bm{\sigma}}:\mathbb{A}:\bm{\sigma}
\end{align}
noticing $\mathbb{C}^{-1}=\frac{1+\nu}{E}\mathbb{I}-\frac{\nu}{E}\bm{\delta}\bm{\delta}$
for this case. Therefore the corresponding topological derivative
$D_{T}y(\Omega,\bm{\xi}_{0})$ has a concrete form 
\begin{equation}
D_{T}y(\Omega,\bm{\xi}_{0})=\tilde{\bm{\sigma}}\left(\bm{\xi}_{0}\right):\mathbb{A}:\bm{\sigma}\left(\bm{\xi}_{0}\right),\label{eq:DT-final}
\end{equation}
where the fourth-order tensor $\mathbb{A}=\frac{2\pi(1-\nu)}{E(7-5\nu)}\left[10(1+\nu)\mathbb{I}-\left(5\nu+1\right)\bm{\delta}\bm{\delta}\right]$.
The evaluation of $D_{T}y(\Omega,\bm{\xi}_{0})$ requires the stress
solution at $\bm{\xi}_{0}$ from both the original problem and the
adjoint problem. In the case that $\bar{\bm{u}}=0$, the latter becomes
self-adjoint and only the solution of Eq. \eqref{eq:BVP} is needed.
The expressions of $\mathbb{A}$ for various cases are summarized
in Table \ref{Table:1}.

\begin{table}
\caption{Displacement solutions on $\partial\omega_{\rho}$ of \eqref{eq:BVP-external}
and tensor $\mathbb{A}$ for various cases}
\label{Table:1} 
\begin{centering}
\begin{tabular}{c|c|c}
\hline 
Isotropic  & Displacement $\textrm{on }\partial\omega_{\rho}$ of Eq. \eqref{eq:BVP-external}  & $\mathbb{A}$\tabularnewline
\hline 
Plane stress  & $\rho\left[\frac{\nu-1}{E}\mathrm{tr}\left(\boldsymbol{\sigma}\left(\boldsymbol{\xi}_{0}\right)\right)\boldsymbol{n}+\frac{3-\nu}{E}\boldsymbol{n}\cdot\boldsymbol{\sigma}\left(\boldsymbol{\xi}_{0}\right)\right]$  & $\frac{\pi}{E}\left[4\mathbb{I}-\boldsymbol{\delta\delta}\right]$\tabularnewline
Plane strain  & $\rho\frac{\left(1+\nu\right)}{E}\left[\left(2\nu-1\right)\mathrm{tr}\left(\boldsymbol{\sigma}\left(\boldsymbol{\xi}_{0}\right)\right)\boldsymbol{n}+\left(3-4\nu\right)\boldsymbol{n}\cdot\boldsymbol{\sigma}\left(\boldsymbol{\xi}_{0}\right)\right]$  & $\frac{\pi\left(1-\nu^{2}\right)}{E}\left[4\mathbb{I}-\boldsymbol{\delta\delta}\right]$\tabularnewline
3D  & $\rho\left[\frac{a-b}{3}\mathrm{tr}\left(\boldsymbol{\sigma}\left(\boldsymbol{\xi}_{0}\right)\right)\boldsymbol{n}+b\boldsymbol{n}\cdot\boldsymbol{\sigma}\left(\boldsymbol{\xi}_{0}\right)\right]\dagger$  & $\frac{2\pi(1-\nu)}{E(7-5\nu)}\left[10(1+\nu)\mathbb{I}-\left(5\nu+1\right)\boldsymbol{\delta\delta}\right]$\tabularnewline
\hline 
\end{tabular}
\par\end{centering}
\centering{}$\dagger a=\frac{1+\nu}{2E},\ b=\frac{2(4-5\nu^{2}-\nu)}{E(7-5\nu)}$,
$\bm{\sigma}\left(\boldsymbol{\xi}_{0}\right)=\mathbb{C}:\bm{\epsilon}\left(\bm{\xi}_{0}\right)$,
where $\boldsymbol{n}$ is the normal of $\partial\omega_{\rho}$ 
\end{table}

\subsection{Topology sensitivity of stochastic moments\label{sec4.2}}

Let $y(\Omega,\mathbf{X})$ be a response function of the linear system
\eqref{eq:BVP} subject to random input $\mathbf{X}$, which can be
uncertain loads, geometry, or material properties. For a point $\bm{\xi}_{0}\in\Omega$,
taking topology derivative of $r$th moments of the response function
$y(\Omega,\mathbf{X})$ and applying the Lebesgue dominated convergence
theorem, which permits the interchange of the differential and integral
operators, yields

\begin{equation}
D_{T}m^{(r)}(\Omega,\bm{\xi}_{0}):=D_{T}\mathbb{E}\left[y^{r}(\Omega,\mathbf{X})\right]\rvert_{\bm{\xi}_{0}}=\int_{\mathbb{R}^{N}}ry^{r-1}(\Omega,\mathbf{x})D_{T}y(\Omega,\mathbf{x},\bm{\xi}_{0})f_{\mathbf{X}}(\mathbf{x})d\mathbf{x}=\mathbb{E}\left[ry^{r-1}(\Omega,\mathbf{X})D_{T}y(\Omega,\mathbf{X},\bm{\xi}_{0})\right],\label{eq:DT-moment}
\end{equation}
that is, the topology derivative is obtained from the expectation
of a product comprised of the response function and its topology derivative.

For simplicity, we denote $D_{T}y(\Omega,\mathbf{X},\bm{\xi}_{0})$
by $z(\Omega,\mathbf{X},\bm{\xi}_{0})$, and construct its $S$-variate,
$m$th-order PDD approximation $\tilde{z}_{S,m}$ as 
\begin{equation}
\tilde{z}_{S,m}(\Omega,\mathbf{X},\bm{\xi}_{0}):=z_{\emptyset}(\Omega,\bm{\xi}_{0})+{\displaystyle \sum_{{\textstyle {\emptyset\ne u\subseteq\{1,\cdots,N\}\atop 1\le|u|\le S}}}}\sum_{{\textstyle {\mathbf{j}_{|u|}\in\mathbb{N}^{|u|}\atop \left\Vert \mathbf{j}_{|u|}\right\Vert _{\infty}\le m}}}\!\!\!\!\!\!D_{u\mathbf{j}_{|u|}}(\Omega,\bm{\xi}_{0})\psi_{u\mathbf{j}_{|u|}}(\mathbf{X}_{u};\Omega),\label{eq:PDD-4-DDT}
\end{equation}

Replacing $y$ and $D_{T}y$ of Eq. \eqref{eq:DT-moment} with their
$S$-variate, $m$th-order PDD approximations $\tilde{y}_{S,m}$ and
$\tilde{z}_{S,m}$, respectively, we have

\begin{equation}
D_{T}\tilde{m}_{S,m}^{(r)}(\Omega,\bm{\xi}_{0})=\mathbb{E}\left[r\tilde{y}_{S,m}^{r-1}(\Omega,\mathbf{X})\tilde{z}_{S,m}(\Omega,\mathbf{X},\bm{\xi}_{0})\right]
\end{equation}
For $r=1,2,3$, employing the zero mean property and orthonormal property
of the PDD basis $\psi_{u\mathbf{j}_{|u|}}(\mathbf{X}_{u};\Omega)$
yields analytical formulations for topology sensitivity of first three
moments

\begin{equation}
D_{T}\tilde{m}_{S,m}^{(1)}(\Omega,\bm{\xi}_{0})=z_{\emptyset}(\Omega,\bm{\xi}_{0}),\label{eq:1st-mom-top-sen}
\end{equation}
\begin{equation}
D_{T}\tilde{m}_{S,m}^{(2)}(\Omega,\bm{\xi}_{0})=2\times\left[y_{\emptyset}(\Omega)z_{\emptyset}(\Omega,\bm{\xi}_{0})+{\displaystyle \sum_{{\textstyle {\emptyset\ne u\subseteq\{1,\cdots,N\}\atop 1\le|u|\le S}}}}\sum_{{\textstyle {\mathbf{j}_{|u|}\in\mathbb{N}^{|u|}\atop ||\mathbf{j}_{|u|}||_{\infty}\le m}}}\!\!\!C_{u\mathbf{j}_{|u|}}(\Omega)D_{u\mathbf{j}_{|u|}}(\Omega,\bm{\xi}_{0})\right],\label{eq:2nd-mom-top-sen}
\end{equation}

\begin{equation}
D_{T}\tilde{m}_{S,m}^{(3)}(\Omega,\bm{\xi}_{0})=3\times\left[z_{\emptyset}(\Omega,\bm{\xi}_{0})\tilde{m}_{S,m}^{(2)}(\Omega)+2y_{\emptyset}(\Omega){\displaystyle \sum_{{\textstyle {\emptyset\ne u\subseteq\{1,\cdots,N\}\atop 1\le|u|\le S}}}}\sum_{{\textstyle {\mathbf{j}_{|u|}\in\mathbb{N}^{|u|}\atop ||\mathbf{j}_{|u|}||_{\infty}\le m}}}\!\!\!C_{u\mathbf{j}_{|u|}}(\Omega)D_{u\mathbf{j}_{|u|}}(\Omega,\bm{\xi}_{0})+T_{k}\right],\label{eq:3rd-mom-top-sen}
\end{equation}

\begin{eqnarray}
 &  & T_{k}={\displaystyle \sum_{{\textstyle {\emptyset\ne u,v,w\subseteq\{1,\cdots,N\}\atop 1\le|u|,|v|,|w|\le S}}}}\sum_{{\textstyle {\mathbf{j}_{|u|},\mathbf{j}_{|v|},\mathbf{j}_{|w|}\in\mathbb{N}^{|u|}\atop ||\mathbf{j}_{|u|}||_{\infty},||\mathbf{j}_{|v|}||_{\infty},||\mathbf{j}_{|w|}||_{\infty}\le m}}}C_{u\mathbf{j}_{|u|}}(\Omega)C_{v\mathbf{j}_{|v|}}(\Omega)D_{w\mathbf{j}_{|w|}}(\Omega,\bm{\xi}_{0})\times\nonumber \\
 &  & \;\;\;\;\;\mathbb{E_{\textrm{\textbf{d}}}}\left[\psi_{u\mathbf{j}_{|u|}}(\mathbf{X}_{u};\Omega)\psi_{v\mathbf{j}_{|v|}}(\mathbf{X}_{v};\Omega)\psi_{wj_{|w|}}(\mathbf{X}_{w};\Omega)\right],\label{21}
\end{eqnarray}
which requires expectations of various products of three random orthonormal
polynomials \citep{ren2013robust}. However, if $\mathbf{X}$ follows
classical distributions such as Gaussian, Exponential, and Uniform
distribution, then the expectations are easily determined from the
properties of univariate Hermite, Laguerre, and Legendre polynomials
\citep{busbridge1948,ren13,rahman2014novel}. For general distributions,
numerical integration methods will apply. \\

\subsection{Topology sensitivity of reliability}

\noindent Using PDD to approximate the performance function $y$,
the Monte Carlo estimate for topology sensitivity of failure probability
is 
\begin{equation}
D_{T}P\left[\mathbf{X}\in\Omega_{F,k}\right]\cong\lim_{\rho\rightarrow0}{\normalcolor {\normalcolor \frac{1}{\rho^{n}}}\lim_{L\rightarrow\infty}}\frac{1}{L}\sum_{l=1}^{L}\left[I_{\tilde{\Omega}_{F,k,\rho}}(\mathbf{x}^{(l)})-I_{\tilde{\Omega}_{F,k}}(\mathbf{x}^{(l)})\right],\label{eq:TDTr}
\end{equation}
where $L$ is the sample size; $x^{(l)}$ is the $l$th realization
of $\mathbf{X}$; $I_{\tilde{\Omega}_{F,k}}$ and $I_{\tilde{\Omega}_{F,k,\rho}}$
are the indicator functions for failure domains $\tilde{\Omega}_{F,k}:=\{\mathbf{x}:\tilde{y}_{k}(\Omega,\mathbf{x})<0\}$
and $\tilde{\Omega}_{F,k,\rho}:=\{\mathbf{x}:\tilde{y}_{k}(\Omega_{\rho},\mathbf{x})<0\}$,
respectively. The PDD approximation of the response function of the
current topology design $\Omega$ is $\tilde{y}_{k}(\Omega,\mathbf{x})$,
while at perturbed design $\Omega_{\rho}$, it is $\tilde{y}_{k}(\Omega_{\rho},\mathbf{x})$.
When $\rho$ takes finite values, Equation \eqref{eq:TDTr} leads
to a finite-difference approximation 
\begin{equation}
D_{T}P\left[\mathbf{X}\in\Omega_{F,k}\right]\cong{\normalcolor {\normalcolor \frac{1}{\rho^{n}}}\lim_{L\rightarrow\infty}}\frac{1}{L}\sum_{l=1}^{L}\left[I_{\tilde{\Omega}_{F,k,\rho}}(\mathbf{x}^{(l)})-I_{\tilde{\Omega}_{F,k}}(\mathbf{x}^{(l)})\right]\label{eq:FD-DT-PF}
\end{equation}
of the topology derivative for reliability. It requires $\tilde{y}_{k}(\Omega_{\rho},\mathbf{X})$,
which is simply obtained from 
\begin{equation}
\tilde{y}_{k}(\Omega_{\rho},\mathbf{X})\cong\tilde{y}_{k}(\Omega,\mathbf{X})+\rho^{n}D_{T}\tilde{y}_{k}(\Omega,\mathbf{X}),\label{eq:perturbed-y-4-DT-PF}
\end{equation}
without additional PDD expansion or FEA involved. This Monte Carlo
estimation entails only two PDD approximations, Eq. \eqref{eq:T-PDD}
for the response function itself and Eq. \eqref{eq:PDD-4-DDT} for
its deterministic topology derivative, both of which are generated
from the same stochastic analysis. Therefore little additional computational
cost is needed to evaluate the topology sensitivity of reliability
once the stochastic analysis is done, facilitating a novel and highly
efficient sensitivity analysis approach for RBTO.

\section{Calculation of PDD Coefficients\label{sec5}}

The expansion coefficients in Eq. \eqref{eq:T-PDD} and Eq. \eqref{eq:PDD-4-DDT}
are defined by $N$-dimensional integrations $y_{\emptyset}(\Omega):=\int_{\mathbb{R}^{N}}y(\mathbf{x})f_{\mathbf{X}}(\mathbf{x})d\mathbf{x}$
and $C_{u\mathbf{j}_{|u|}}(\Omega):=\int_{\mathbb{R}^{N}}y(\mathbf{x})\psi_{u\mathbf{j}_{|u|}}(\mathbf{X}_{u};\Omega)f_{\mathbf{X}}(\mathbf{x})d\mathbf{x}$
etc. For large $N$, direct numerical integration is often prohibitive,
especially when FEA is involved in the Gauss point evaluation. Instead,
we will use the dimension-reduction method \citep{xu04,rahman2004,xu05},
which entails multiple low-dimensional integrations as an effective
replacement of a single $N$-dimensional integration.

Let $\mathbf{c}=(c_{1},\cdots,c_{N})^{T}\in\mathbb{R}^{N}$, which
is commonly adopted as the mean of $\mathbf{X}$, be a reference point,
and $y(\mathbf{x}_{v},\mathbf{c}_{-v})$ represent an $|v|$-variate
referential dimensional decomposition (RDD) component function of
$y(\mathbf{X})$, where $v\subseteq\{1,\text{\ensuremath{\cdots}},N\}$
and $-v=\left\{ 1,\text{\ensuremath{\cdots}},N\right\} \backslash v$.
Given a positive integer $S\le R\le N$, when $y(\mathbf{x})$ in
the above $N$-dimensional integration is replaced by its $R$-variate
RDD approximation, the coefficients are estimated from\citep{xu04}

\begin{eqnarray}
y_{\emptyset}(\Omega) & \cong & {\displaystyle \sum_{i=0}^{R}}(-1)^{i}{N-R+i-1 \choose i}\sum_{{\textstyle {v\subseteq\{1,\cdots,N\}\atop |v|=R-i}}}\!\int_{\mathbb{R}^{|v|}}y(\mathbf{x}_{v},\mathbf{c}_{-v})f_{\mathbf{X}_{v}}(\mathbf{x}_{v})d\mathbf{x}_{v}\label{eq:y0-DR}
\end{eqnarray}

\begin{eqnarray}
C_{u\mathbf{j}_{|u|}}(\Omega) & \cong & {\displaystyle \sum_{i=0}^{R}}(-1)^{i}{N-R+i-1 \choose i}\sum_{{\textstyle {v\subseteq\{1,\cdots,N\}\atop |v|=R-i,u\subseteq v}}}\!\int_{\mathbb{R}^{|v|}}y(\mathbf{x}_{v},\mathbf{c}_{-v})\psi_{u\mathbf{j}_{|u|}}(\mathbf{x}_{u};\Omega)f_{\mathbf{X}_{v}}(\mathbf{x}_{v})d\mathbf{x}_{v}\label{eq:cuju-DR}
\end{eqnarray}
entailing at most $R$-dimensional integrations. For each integration
involved, the Gauss quadrature rule applies. For engineering problems,
the evaluation of Gauss points often relies on FEA.\textcolor{black}{{}
For instance, each FEA with $\mathbf{X}$ realized at certain gauss
point supplies response function value for that Gauss point. Whereas
to approximate the coefficients for }the topology sensitivity $D_{T}y(\Omega,\mathbf{X},\bm{\xi}_{0})$
or $z(\Omega,\mathbf{X},\bm{\xi}_{0})$ in section \ref{sec4.2},
\textcolor{black}{each FEA provides stress results for Eq. \ref{eq:DT-final}
and further produces }$z$ values at the corresponding Gauss point.
\textcolor{black}{Nonetheless the reduced integration is }significantly
more efficient than performing one $N$-dimensional integration owing
to a much fewer number of Gauss points required by the former, particularly
when $R\ll N$. Moreover, it \textcolor{black}{facilitates the calculation
of coefficients approaching their exact value as} $R\rightarrow N$.
In addition, the same set of Gauss points thus the same set of FEAs
will be reused for the evaluation of coefficients in Eq. \eqref{eq:PDD-4-DDT},
rendering a significantly efficient framework for stochastic topology
sensitivity analysis.

\section{Numerical Examples\label{sec6}}

In this section, two new benchmark examples are developed for the
analytical or semi-analytical solution of moments and reliability
and their topology sensitivities. The first one involves two random
variables and renders analytical expression for both stochastic quantities
and their topology sensitivities of compliance. The second one contains
53 random variables to test the accuracy and efficiency of the proposed
method for high dimensional problems by developing corresponding analytical
and semi-analytical solutions. The third example is a three-dimensional
bracket, whose topology has already been optimized, illustrating a
practical application of the proposed method. In all examples, orthonormal
polynomials and associated Gauss quadrature rules consistent with
the probability distributions of input variables, including classical
forms, if they exist, were employed. No unit for length, force, and
Young's modulus is specified in all examples for simplicity while
permitting any consistent unit system for the results.

\subsection{A round disk subject to a uniform pressure}

\begin{figure}
\centering{}\includegraphics[scale=1.8]{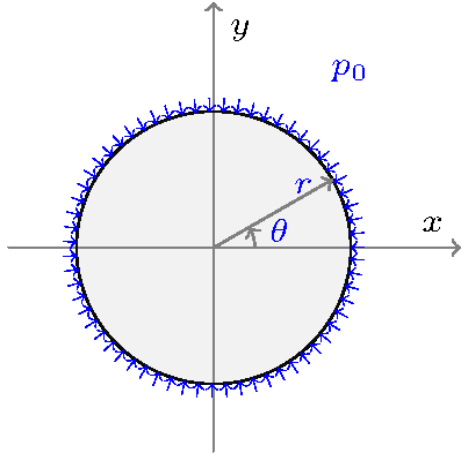}\caption{A round disk subject to a uniform pressure}
\label{Fig:Disk-pressure} 
\end{figure}

Assuming the plane stress state, consider a round disk $\Omega=\left\{ \left(r,\theta\right):r\le1,\theta\in[0,2\pi)\right\} $
subject to a uniform pressure $p_{0}$ as shown in Fig. \ref{Fig:Disk-pressure},
where $\left(r,\theta\right)$ is the polar coordinate system with
its origin locating at the center of the disk. The Young's module
$E$ and pressure $p_{0}$ are random variables. The Poisson's ratio
$\nu=0.2$, and is deterministic. 

Assume $E$ follows inverse uniform distribution on $\left[2,4\right]$
with the probability density function (PDF) 
\begin{equation}
f_{E}(x_{E})=4x_{E}^{-2}
\end{equation}
and $P_{0}$ follows uniform distribution on $\left[1,2\right]$.
For this particular problem, the exact compliance is readily available,
it is 
\begin{equation}
y\left(\Omega\right)=2\pi\frac{1-\nu}{E}p_{0}^{2}.\label{eq:analytical-compliance-uniform-pressure}
\end{equation}
The exact PDF of the compliance for this particular problem is found
as 
\begin{equation}
f_{Y}\left(y\right)=\begin{cases}
\frac{1}{\pi\left(1-\nu\right)}\left(2-\left(\frac{y}{2\pi\left(1-\nu\right)}\right)^{-\frac{1}{2}}\right) & \frac{\pi\left(1-\nu\right)}{2}\le y<\pi\left(1-\nu\right)\\
\frac{1}{\pi\left(1-\nu\right)}\left(\frac{y}{2\pi\left(1-\nu\right)}\right)^{-\frac{1}{2}}\left(\sqrt{2}-1\right) & \pi\left(1-\nu\right)\le y<2\pi\left(1-\nu\right)\\
\frac{1}{\pi\left(1-\nu\right)}\left(\sqrt{2}\left(\frac{y}{2\pi\left(1-\nu\right)}\right)^{-\frac{1}{2}}-1\right) & 2\pi\left(1-\nu\right)\le y<4\pi\left(1-\nu\right)
\end{cases}.\label{eq:pdf-compliance}
\end{equation}
Moreover, the analytical expression of the first three moments of
compliance are summarized in the Table \ref{Tab:uniform-pressure-disk-mom}

\begin{table}
\raggedright{}\centering{}\caption{Analytical solutions, numerical results, and relative errors: moments}
\label{Tab:uniform-pressure-disk-mom}%
\begin{tabular}{>{\raggedright}p{2.7cm}|c|>{\centering}m{1.5cm}|c|>{\centering}m{1.5cm}|c|>{\centering}m{1.35cm}}
\hline 
\multirow{1}{2.7cm}{} & \multicolumn{2}{c|}{$m^{(1)}$} & \multicolumn{2}{c|}{$m^{(2)}$} & \multicolumn{2}{c}{$m^{(3)}$}\tabularnewline
\cline{2-7} \cline{3-7} \cline{4-7} \cline{5-7} \cline{6-7} \cline{7-7} 
 & values  & Relative Error (\%)  & values & Relative Error (\%)  & values  & Relative Error (\%)\tabularnewline
\hline 
PDD $S=1,\ m=1$  & 4.387142651 & 0.252  & 22.31764876 & 2.308  & 124.853276 & 7.523\tabularnewline
PDD $S=1,\ m=2$  & 4.392067155 & 0.140  & 22.43673535 & 1.786  & 127.987972 & 5.201\tabularnewline
PDD $S=1,\ m=3$  & 4.392213984 & 0.137  & 22.44094545 & 1.768  & 128.078790 & 5.134\tabularnewline
PDD $S=2,\ m=1$  & 4.392955528 & 0.120  & 22.69742049 & 0.645  & 131.939662 & 2.274\tabularnewline
PDD $S=2,\ m=2$  & 4.398062552 & 0.004  & 22.83944821 & 0.024  & 136.197391 & 0.879\tabularnewline
PDD $S=2,\ m=3$  & 4.398214737 & $3.4\times10^{-4}$  & 22.84455042 & $1.3\times10^{-3}$  & 136.337101 & 0.983\tabularnewline
\hline 
Analytical  & \multicolumn{2}{c|}{$\frac{7\pi}{4}\left(1-\nu\right)$} & \multicolumn{2}{c|}{$\frac{217\pi^{2}}{60}\left(1-\nu\right)^{2}$} & \multicolumn{2}{c}{$\frac{1905\pi^{3}}{224}\left(1-\nu\right)^{3}$}\tabularnewline
\hline 
\end{tabular}
\end{table}

To calculate the analytical topology sensitivity of moments and failure
probability at the center $\boldsymbol{\xi}_{0}$, another analytical
solution for the perforated domain with a tiny hole at the center
is needed. It reads 
\begin{equation}
y\left(\Omega_{\rho}\right)=\frac{2\pi p_{o}^{2}}{E\left(1-\rho^{2}\right)}\left[\left(1+\nu\right)\rho^{2}+\left(1-\nu\right)\right],
\end{equation}
which can be derived based on the \textit{Lamé's strain potential}
$C\ln\frac{r}{K}$ with undetermined constants $C$ and $K$ via the
displacement method. The deterministic topology derivative $D_{T}y$
by definition is 
\begin{equation}
D_{T}y\text{\ensuremath{\left(\Omega,\boldsymbol{\xi}_{0}\right)}}=\lim_{\epsilon\rightarrow0}\frac{y\left(\Omega_{\rho}\right)-y\left(\Omega\right)}{\rho^{2}}=\frac{4\pi p_{0}^{2}}{E}.\label{eq:anlytical-DT-compliance-uniformPresser-center}
\end{equation}
Together with Eqs. \eqref{eq:DT-moment} and \eqref{eq:pdf-compliance},
the analytical expressions of topology sensitivity for the first three
moments can be determined, and are listed in Table \ref{Tab:uniform-pressure-disk-mom-sen}.

\begin{table}
\centering{}\caption{Analytical solutions, numerical results, and relative errors: sensitivity
of moments at $\bm{\xi}_{0}=\left(0,0\right)$}
\label{Tab:uniform-pressure-disk-mom-sen}%
\begin{tabular}{>{\centering}p{3.1cm}|>{\centering}p{2.1cm}|>{\centering}m{1.5cm}|>{\centering}p{2.1cm}|>{\centering}m{1.5cm}|>{\centering}p{2.1cm}|>{\centering}m{1.5cm}}
\hline 
\multirow{1}{3.1cm}{} & \multicolumn{2}{c|}{$D_{T}m^{(1)}(\Omega,\bm{\xi}_{0})$} & \multicolumn{2}{c|}{$D_{T}m^{(2)}(\Omega,\bm{\xi}_{0})$} & \multicolumn{2}{c}{$D_{T}m^{(3)}(\Omega,\bm{\xi}_{0})$}\tabularnewline
\cline{2-7} \cline{3-7} \cline{4-7} \cline{5-7} \cline{6-7} \cline{7-7} 
 & values  & Relative Error (\%)  & values  & Relative Error (\%)  & values  & Relative Error (\%)\tabularnewline
\hline 
PDD $S=1,\ m=1$  & 10.96790142  & 0.252  & 111.5886904  & 2.307  & 936.4032737  & 7.522\tabularnewline
PDD $S=1,\ m=2$  & 10.98019122  & 0.140  & 112.1839012  & 1.786  & 959.9116087  & 5.201\tabularnewline
PDD $S=1,\ m=3$  & 10.98056234  & 0.137  & 112.2050067  & 1.768  & 960.5932965  & 5.134\tabularnewline
PDD $S=2,\ m=1$  & 10.98243457  & 0.120  & 113.4875794  & 0.645  & 989.5516625  & 2.274\tabularnewline
PDD $S=2,\ m=2$  & 10.99517968  & 0.004  & 114.1974516  & 0.023  & 1021.481992  & 0.879\tabularnewline
PDD $S=2,\ m=3$  & 10.99556439  & $9.0\times10^{-5}$  & 114.2230294  & $1.0\times10^{-3}$  & 1022.530638  & 0.983\tabularnewline
\hline 
Analytical  & \multicolumn{2}{c|}{$\frac{7\pi}{2}$} & \multicolumn{2}{c|}{$\frac{217\pi^{2}}{15}\left(1-\nu\right)$} & \multicolumn{2}{c}{$\frac{5715\pi^{3}}{112}\left(1-\nu\right)^{2}$}\tabularnewline
\hline 
\end{tabular}
\end{table}

The finite element model employed in the proposed method consists
of 404800 quadrilateral and 1600 triangular elements. The displacement
$u_{\theta}$ at $\left(1,0\right)$, $\left(1,\frac{\pi}{2}\right),$
and $\left(1,\frac{3\pi}{2}\right),$ are specified as \textit{zero}
to make the FEA model well-posed and keep the same solution of stress,
strain, and compliance in Fig. \ref{Fig:Disk-pressure}. Table \ref{Tab:uniform-pressure-disk-mom}
displays the approximate moments of the compliance, committed by the
proposed univariate $(S=1)$ and bivariate $(S=2)$ PDD for $m=1,2,3$.
Relative errors, defined as the ratio of the absolute error to the
exact value, are also presented. For the first moments, the errors
range from $3.4\times10^{-4}$ to $0.252$ percent. When the order
of moments increases, the errors show an uptrend as expected due to
the accumulation of approximation errors, but still maintains good
levels, $1.3\times10^{-3}$ to $2.308$ percent for the second moments
and $0.983$ to $7.523$ percent for the third moments.

Table \ref{Tab:uniform-pressure-disk-mom-sen} presents the approximate
topology sensitivity of the center point and their relative errors
for the first three moments. For the same set of $S$ and $m$ values,
the relative errors of topology sensitivity are almost identical with
the ones of moments in Table \ref{Tab:uniform-pressure-disk-mom}.
It seems unusual since for many methods the numerical estimation of
stochastic sensitivity is often less accurate than the estimation
of the function itself. However, the proposed method dovetails the
deterministic topology derivative $D_{T}y$ as shown in Eq. \eqref{eq:DT-moment}
and the nonlinearity and interactive effects in $D_{T}y$ are often
similar with the response $y$ as shown in Eqs \eqref{eq:analytical-compliance-uniform-pressure}
and \eqref{eq:anlytical-DT-compliance-uniformPresser-center}, which
lead to similar or identical relative errors in the sensitivity of
moments. The errors from the propose method drop as $m$ and $S$
increase as expected for both moments and their topology sensitivities.

Analytical expressions and numerical results of failure probabilities
and their topology sensitivity are presented in Tables \ref{Tab:reliability-sensitivity-at-7.0-uniformPressure}
and \ref{Tab:reliability-sensitivity-at-7.5-uniformPressure} for
two limit-state values $7.0$ and $7.5$, respectively. The numerical
estimations of failure probability by the proposed method are evaluated
via Eq. \eqref{Eq:pfmcs} using the embedded MCS, whereas their topology
sensitivities are calculated based on Eqs. \eqref{eq:FD-DT-PF} and
\eqref{eq:perturbed-y-4-DT-PF} with a finite $\rho$ value of $0.05$.
The sample size for both is $L=10^{6}$. The total number of FEA simulations
for various combinations of the truncation parameters $S=1,2$ and
$m=1,2,3$ are listed in the Table \ref{Tab:reliability-sensitivity-at-7.5-uniformPressure}.
It is worthy to note that one set of FEAs generate the associated
PDD approximations for both the response function and its deterministic
topology derivatives at the same time. In addition, the two PDD approximations
deliver stochastic analyses and stochastic topology sensitivity analyses,
generating moments, reliabilities, and their topology sensitivities
without additional FEAs. The errors of failure probability and its
sensitivity by the linear ($m=1$) univariate ($S=1$) PDD are relatively
large, but it requires only 5 FEAs. But the errors drop significantly
as $S$ and/or $m$ increases. For instance, the errors of failure
probability become less than one percent for $S=2,\ m=2,3$, requiring
15 and 25 FEAs respectively. Similar trends are observed in their
topology sensitivity. Comparing results for $\bar{y}=7.0$ and $\bar{y}=7.5$,
the errors of failure probability increase as expected when the limit
state values move away from the mean of the response function. Further
developments address this problem in our future work.

\begin{table}
\centering{}\caption{comparison between analytical solution and numerical results: reliability
and its sensitivity for $\rho=0.05,\bar{y}=7.0$ at $\bm{\xi}_{0}=\left(0,0\right)$}
\label{Tab:reliability-sensitivity-at-7.0-uniformPressure}%
\begin{tabular}{c|c|>{\centering}m{1.5cm}|c|>{\centering}m{1.5cm}}
\hline 
\multirow{1}{*}{} & \multicolumn{2}{c|}{$P_{F}:=P\left(y\ge\bar{y}\right)$, $\bar{y}=7.0$} & \multicolumn{2}{c}{$D_{T}P_{F}(\Omega,\bm{\xi}_{0})$}\tabularnewline
\cline{2-5} \cline{3-5} \cline{4-5} \cline{5-5} 
 & values  & Relative Error (\%)  & values  & Relative Error (\%)\tabularnewline
\hline 
PDD $S=1,\ m=1$  & $0.72325\times10^{-1}$  & 34.028  & $1.9776$  & 43.154\tabularnewline
PDD $S=1,\ m=2$  & $0.85563\times10^{-1}$  & 21.953  & $1.6684$  & 20.772\tabularnewline
PDD $S=1,\ m=3$  & $0.85065\times10^{-1}$  & 22.407  & $1.6516$  & 19.556\tabularnewline
PDD $S=2,\ m=1$  & $0.105089$  & 4.142  & $1.6744$  & 21.206\tabularnewline
PDD $S=2,\ m=2$  & $0.110683$  & 0.9602  & 1.3936  & 0.8797\tabularnewline
PDD $S=2,\ m=3$  & $0.109681$  & 0.04622  & 1.4284  & 3.3988\tabularnewline
\hline 
Analytical($\rho\rightarrow0$)  & \multicolumn{2}{c|}{$1-\frac{4\sqrt{5.6\pi}-2.4\pi-7}{0.8\pi}$} & \multicolumn{2}{c}{$\frac{4\sqrt{5.6\pi}-14}{0.64\pi}$}\tabularnewline
\hline 
\end{tabular}
\end{table}

\begin{table}
\centering{}\caption{comparison between analytical solution and numerical results: reliability
and its sensitivity for $\rho=0.05,\bar{y}=7.5$ at $\bm{\xi}_{0}=\left(0,0\right)$}
\label{Tab:reliability-sensitivity-at-7.5-uniformPressure}%
\begin{tabular}{c|c|>{\centering}m{1.5cm}|c|>{\centering}m{1.5cm}|c}
\hline 
\multirow{1}{*}{} & \multicolumn{2}{c|}{$P_{F}:=P\left(y\ge\bar{y}\right)$, $\bar{y}=7.5$} & \multicolumn{2}{c|}{$D_{T}P_{F}(\Omega,\bm{\xi}_{0})$ } & \multirow{2}{*}{\# of FEA}\tabularnewline
\cline{2-5} \cline{3-5} \cline{4-5} \cline{5-5} 
 & values  & Relative Error (\%)  & values  & Relative Error (\%)  & \tabularnewline
\hline 
PDD $S=1,\ m=1$  & $0.26411\times10^{-1}$  & 64.440  & 1.3572  & 15.313  & 5\tabularnewline
PDD $S=1,\ m=2$  & $0.44418\times10^{-1}$  & 40.196  & 1.3540  & 15.041  & 5\tabularnewline
PDD $S=1,\ m=3$  & $0.44335\times10^{-1}$  & 40.307  & 1.3052  & 10.895  & 9\tabularnewline
PDD $S=2,\ m=1$  & $0.62140\times10^{-1}$  & 16.335  & 1.4212  & 20.751  & 9\tabularnewline
PDD $S=2,\ m=2$  & $0.749340\times10^{-1}$  & 0.8911  & 1.1672  & 0.8297  & 15\tabularnewline
PDD $S=2,\ m=3$  & $0.741870\times10^{-1}$  & 0.1147  & 1.1748  & 0.1840  & 25\tabularnewline
\hline 
Analytical($\rho\rightarrow0$)  & \multicolumn{2}{c|}{$1-\frac{4\sqrt{6\pi}-2.4\pi-7.5}{0.8\pi}$} & \multicolumn{2}{c|}{$\frac{4\sqrt{6\pi}-15}{0.64\pi}$} & NA\tabularnewline
\hline 
\end{tabular}
\end{table}

\subsection{A 53-random-variable example: the round disk subject to pressure
in terms of trigonometric functions}

\begin{figure}
\centering{}\includegraphics[scale=1.8]{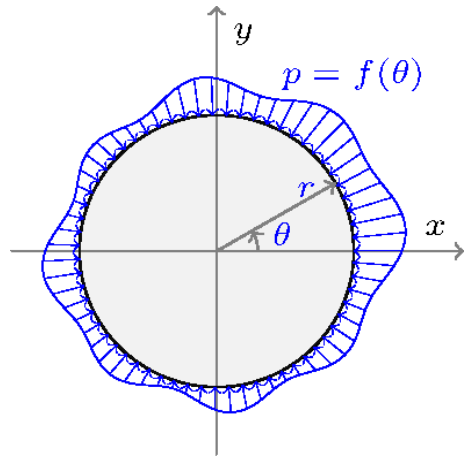}\caption{A round disk subject to a complex pressure}
\label{Fig:Disk-complex-pressure} 
\end{figure}

Consider the same round disk in last example but subject to a more
complex pressure as shown in Fig. \ref{Fig:Disk-complex-pressure},
where the pressure function 
\begin{equation}
f\left(\theta\right)=D_{0}+\sum_{k=1}^{K}\left(D_{k}\cos\left(k+1\right)\theta+E_{k}\sin\left(k+1\right)\theta\right)\label{eq:trigonometric-pressure-function}
\end{equation}
accommodating $2K+1$ random variables $D_{k},\ k=0,\cdots,K$ and
$E_{k},\ k=1,\cdots,K$. 

\subsubsection{Analytical solutions}

Employing the \textit{Taylor series expansion} of holomorphic functions
in a simply-connected domain and \textit{Goursat formula} \citep{muskhelishvili1963some},
the analytical solution for compliance of the disk subject to the
above pressure is found in the form of 
\begin{equation}
y\left(\Omega\right)=\frac{2D_{0}^{2}\pi\left(1-\nu\right)}{E}+\sum_{k=1}^{K}\frac{\left(D_{k}^{2}+E_{k}^{2}\right)\pi\left(\nu+2k+1\right)}{k\left(k+2\right)E}.\label{eq:disk-solution}
\end{equation}
The solution \eqref{eq:disk-solution} is general and applicable for
the pressure function \eqref{eq:trigonometric-pressure-function}
for any positive integer $K$.

Now consider perforating a tiny hole of radius $\rho$ in the center
of the disk. Its compliance, subject to the same pressure function
\eqref{eq:trigonometric-pressure-function}, is found as follows 
\begin{equation}
y\left(\Omega_{\rho}\right)=\frac{2D_{0}^{2}\pi\left[\rho^{2}\left(1+\nu\right)+\left(1-\nu\right)\right]}{E\left(1-\rho^{2}\right)}+\sum_{k=1}^{K}\frac{A_{k}B_{k}}{C_{k}F_{k}},\label{eq:ring-solution}
\end{equation}
where 
\[
A_{k}=\left(D_{k}^{2}+E_{k}^{2}\right)\pi
\]
\[
B_{k}=\rho^{2k}\left(k+2\right)\left[k\nu-\left(3k+2\right)-\left(k\nu+k+2\right)\rho^{2}\right]+\left[\left(\nu-2k-3\right)\rho^{2\left(k+2\right)}-\left(\nu+2k+1\right)\right]\sum_{j=0}^{k-1}\rho^{2j}
\]

\[
C_{k}=k\left(k+2\right)E
\]

\[
F_{k}=k\left(k+2\right)\rho^{2k}\left(1-\rho^{2}\right)+\left(\rho^{2\left(k+2\right)}-1\right)\sum_{j=0}^{k-1}\rho^{2j}.
\]
which requires \textit{Laurent series expansion }of holomorphic functions
in a double-connected region.

Employing Eqs. \eqref{eq:disk-solution} and \eqref{eq:ring-solution},
the analytical expression of the deterministic topology derivative
at the center reads 
\begin{equation}
D_{T}y\left(\Omega,\bm{\xi}_{0}\right)=\lim_{\rho\rightarrow0}\frac{y\left(\Omega_{\rho}\right)-y\left(\Omega\right)}{\rho^{2}}=\frac{4\pi\left(D_{0}^{2}+2D_{1}^{2}+2E_{1}^{2}\right)}{E},\label{eq:DT-compliance-center}
\end{equation}
indicating that at the center of the disk the topology derivative
of compliance is merely related to Young's modulus $E$ and three
parameters $D_{0}$, $D_{1}$, $E_{1}$ in the pressure function.

The exact topological sensitivities of moments at the center are derived
from 
\begin{equation}
D_{T}m^{(r)}(\Omega,\bm{\xi}_{0})=\int_{\mathbb{R}^{N}}ry^{r-1}(\Omega,\mathbf{X})D_{T}y(\Omega,\mathbf{X},\bm{\xi}_{0})f_{\mathbf{X}}(\mathbf{x})d\mathbf{x},\label{eq:DT-moment-Exact}
\end{equation}
employing Eqs. \eqref{eq:disk-solution} and \eqref{eq:DT-compliance-center}.
Generally, Eq. \eqref{eq:DT-moment-Exact} admits any proper distributions
for the $2K+1$ random variables.

\subsubsection{Benchmarks}

Let $K=25$, random variables $D_{k},\ k=0,\cdots,25$ and $E_{k},\ k=1,\cdots,25$
follow four-parameter Beta distributions with mean value $\mu_{D_{k}}=k+1$,
$\mu_{E_{k}}=k+1$, and coefficient of variance (CV) be $0.1$ for
all $D_{k}$ and $E_{k}$. Two isotropic elastic material constants
also follow four-parameter Beta distributions, where Young's modulus
$E$ has a mean value of $10^{6}$ and CV of $0.1$, the Poisson's
ratio $\nu$ has a mean value of 0.2 and CV of $0.01$. The support
of each Beta variable is $\left[\mu-3\sigma,\mu+3\sigma\right]$,
where $\mu$ and $\sigma$ here denote mean and standard deviation
of the corresponding variable.

The exact solutions of the first three moments of the compliance,
obtained based on the analytical solution \eqref{eq:disk-solution},
are exhibited in Table \ref{Tab:complex-pressure-disk-mom-coarse}.
For the finite element model used in the proposed method, two types
of mesh are adopted: 1) coarse mesh (24800 quadrilateral and 400 triangular
elements), and 2) fine mesh (404800 quadrilateral and 1600 triangular
elements), as shown in Tables \ref{Tab:complex-pressure-disk-mom-coarse}
and \ref{Tab:complex-pressure-disk-mom-fine}. The displacement $u_{\theta}$
at $\left(1,0\right)$, $\left(1,\frac{\pi}{2}\right),$ and $\left(1,\frac{3\pi}{2}\right),$
are specified as \textit{zero} to make the FEA model well-posed and
meanwhile keep the compliance unchanged. For the results by the coarse
mesh, the relative errors of the first moment by the proposed method
with various truncations range from $1.056$ to $1.146$ percent.
When the order of moments increases, the relative errors rise, for
instance, to $2.308$-$2.532$ percent for the second moment and to
$3.749$-$4.183$ percent for the third moment. This trend is foreseeable
since the moment calculation accumulates the error of the approximated
response function when its order increases. Checking any particular
moment in Table \ref{Tab:complex-pressure-disk-mom-coarse}, the prevailing
trend of the relative errors is down when increasing truncation parameters
$S$ and $m$, but it is insignificant. The reason as disclosed in
the later discussion is that the error introduced by FEA approximations
is dominant comparing to the error of the PDD approximation. Nonetheless,
roughly $1.1$ percent error for $m^{(1)}$, $2.4$ percent error
for $m^{(2)}$, and $4.0$ percent error for $m^{(3)}$ are highly
satisfactory for stochastic moment analysis using the coarse mesh.
When employing the fine mesh, the relative errors of all three moments
plummet approximately by half for every combination of truncation
parameters as shown in Table \ref{Tab:complex-pressure-disk-mom-fine},
which indicates the error from FEA may dominate the error of PDD approximations.
The relative errors for $m^{(1)}$, $m^{(2)}$, and $m^{(3)}$ by
the proposed method using the fine mesh are merely $0.4$, $1.1$,
and $2.0$ percent, respectively.

The topology sensitivities for the first three moments of compliance
are examined at the center point $\boldsymbol{\xi}_{0}=\left(0,0\right)$,
indicating the change ratio of the three moments after perforating
a tiny hole at $\boldsymbol{\xi}_{0}$. Their exact solutions are
unveiled in Tables \ref{Tab:complex-pressure-disk-mom-sen-coarse}
and \ref{Tab:complex-pressure-disk-mom-sen-fine}. The proposed method
is implemented in all combinations of $S=1,2$ and $m=1,2,3$ for
various PDD truncations and the corresponding results by coarse and
fine mesh are listed in Tables \ref{Tab:complex-pressure-disk-mom-sen-coarse}
and \ref{Tab:complex-pressure-disk-mom-sen-fine}, respectively. It
is noteworthy that the proposed method for topology sensitivity of
moments roots in Eqs. \eqref{eq:1st-mom-top-sen}-\eqref{eq:3rd-mom-top-sen},
which dovetail PDD approximation of the deterministic topology derivative
of the response. The ranges of relative errors for the topology sensitivities
by the proposed method are $\left[0.066,0.422\right]$, $\left[1.168,1.800\right]$,
and $\left[2.614,3.444\right]$ when using the coarse mesh. Whereas
using the fine mesh, they are $\left[0.193,0.389\right]$, $\left[0.808,1.058\right]$,
and $\left[1.627,2.011\right]$, respectively, showing significant
drops especially in the errors of $D_{T}m^{(2)}(\Omega,\bm{\xi}_{0})$
and $D_{T}m^{(3)}(\Omega,\bm{\xi}_{0})$. Tables \ref{Tab:complex-pressure-disk-mom-coarse}-\ref{Tab:complex-pressure-disk-mom-sen-fine}
demonstrate that the proposed method is capable of performing highly
accurate moment analysis as well as their topology sensitivity analysis.
By comparing results from two mesh cases, it can be inferred that
a significant portion of errors come from FEA, conjointly evincing
the accuracy of the proposed method. Moreover, sensitivity analyses
not limited to topology sensitivity analyses of a generic response
function are often less accurate than the evaluation of the function
itself. However, comparing Table \ref{Tab:complex-pressure-disk-mom-coarse}
with Table \ref{Tab:complex-pressure-disk-mom-sen-coarse}, or Table
\ref{Tab:complex-pressure-disk-mom-fine} with Table \ref{Tab:complex-pressure-disk-mom-sen-fine},
it shows that for the same mesh case and the same set of $S$ and
$m$ the topology sensitivity is surprisingly more accurate than the
moment analysis itself. For instance,$1.800$ percent error for $D_{T}m^{(2)}(\Omega,\bm{\xi}_{0})$
is less than $2.532$ percent error for $m^{(2)}$ itself in the case
of coarse mesh, $S=1$, and $m=1$. The remarkable more accuracy of
sensitivity seems occasional and rare, however, it is reasonable for
the proposed framework due to the deterministic topology embedded
in Eqs. \eqref{eq:DT-moment}-\eqref{eq:3rd-mom-top-sen}. Scrutinizing
the definition of the $r$th moment $m^{(r)}(\Omega):=\mathbb{E}[y^{r}(\Omega,\mathbf{X})]$
and its topology sensitivity Eq. \eqref{eq:DT-moment}, a major difference
between them is the replacement of $y$ by $D_{T}y$ in the topology
sensitivity. When the nonlinearity and interaction structure of $D_{T}y$
is equal or simpler than ones of $y$, for the same set of truncation
parameter $S$ and $m$, the topology sensitivity of moments calculated
by the proposed method is bound to be equally or more accurate than
the moments itself. The deterministic topology derivative at the center
for this example is shown in Eq. \eqref{eq:DT-compliance-center},
which is obviously simpler than the compliance itself as shown in
Eq. \eqref{eq:disk-solution}. The structure of the proposed method
in Eqs. \eqref{eq:DT-moment}-\eqref{eq:3rd-mom-top-sen} well explains
the observation that topology sensitivity is more accurate than the
moment itself and also demonstrates another advantage of the new method.

\begin{table}
\centering{}\caption{Exact solutions, numerical results, and relative errors for moments
- coarse mesh}
\label{Tab:complex-pressure-disk-mom-coarse}%
\begin{tabular}{c|c|>{\centering}m{1.5cm}|c|>{\centering}m{1.5cm}|c|>{\centering}m{1.5cm}}
\hline 
\multirow{1}{*}{} & \multicolumn{2}{c|}{$m^{(1)}$} & \multicolumn{2}{c|}{$m^{(2)}$} & \multicolumn{2}{c}{$m^{(3)}$}\tabularnewline
\cline{2-7} \cline{3-7} \cline{4-7} \cline{5-7} \cline{6-7} \cline{7-7} 
 & values  & Relative Error (\%)  & values  & Relative Error (\%)  & values  & Relative Error (\%)\tabularnewline
\hline 
PDD $S=1,\ m=1$  & 4.35037888E-3  & 1.146  & 1.90933140E-5  & 2.532  & 8.452068041E-8  & 4.183\tabularnewline
PDD $S=1,\ m=2$  & 4.35088784E-3  & 1.134  & 1.91037093E-5  & 2.479  & 8.46574593E-8  & 4.028\tabularnewline
PDD $S=1,\ m=3$  & 4.35081912E-3  & 1.136  & 1.91032666E-5  & 2.481  & 8.46571311E-8  & 4.029\tabularnewline
PDD $S=2,\ m=1$  & 4.35332715E-3  & 1.079  & 1.91218647E-5  & 2.386  & 8.47307023E-8  & 3.945\tabularnewline
PDD $S=2,\ m=2$  & 4.35171366E-3  & 1.116  & 1.91138791E-5  & 2.427  & 8.47479823E-8  & 3.926\tabularnewline
PDD $S=2,\ m=3$  & 4.35436004E-3  & 1.056  & 1.91370811E-5  & 2.308  & 8.49035726E-8  & 3.749\tabularnewline
\hline 
Exact  & \multicolumn{2}{c|}{4.400814209E-3} & \multicolumn{2}{c|}{1.958928121E-5} & \multicolumn{2}{c}{8.821066188E-8}\tabularnewline
\hline 
\end{tabular}
\end{table}

\begin{table}
\centering{}\caption{Exact solutions, numerical results, and relative errors for moments
- fine mesh}
\label{Tab:complex-pressure-disk-mom-fine}%
\begin{tabular}{c|c|>{\centering}p{1.5cm}|c|>{\centering}p{1.5cm}|c|>{\centering}p{1.5cm}}
\hline 
\multirow{1}{*}{} & \multicolumn{2}{c|}{$m^{(1)}$} & \multicolumn{2}{c|}{$m^{(2)}$} & \multicolumn{2}{c}{$m^{(3)}$}\tabularnewline
\cline{2-7} \cline{3-7} \cline{4-7} \cline{5-7} \cline{6-7} \cline{7-7} 
 & values  & Relative Error (\%)  & values  & Relative Error (\%)  & values  & Relative Error (\%)\tabularnewline
\hline 
PDD $S=1,\ m=1$  & 4.38180775E-3  & 0.432  & 1.93702207E-5  & 1.118  & 8.63662287E-8  & 2.091\tabularnewline
PDD $S=1,\ m=2$  & 4.38225397E-3  & 0.422  & 1.93801836E-5  & 1.067  & 8.65021250E-8  & 1.937\tabularnewline
PDD $S=1,\ m=3$  & 4.38226696E-3  & 0.421  & 1.93804554E-5  & 1.066  & 8.65065744E-8  & 1.932\tabularnewline
PDD $S=2,\ m=1$  & 4.38201078E-3  & 0.427  & 1.93749344E-5  & 1.094  & 8.64199812E-8  & 2.030\tabularnewline
PDD $S=2,\ m=2$  & 4.38253889E-3  & 0.415  & 1.93857124E-5  & 1.039  & 8.65628909E-8  & 1.868\tabularnewline
PDD $S=2,\ m=3$  & 4.38229842E-3  & 0.421  & 1.93837562E-5  & 1.049  & 8.65525631E-8  & 1.880\tabularnewline
\hline 
Exact  & \multicolumn{2}{c|}{4.400814209E-3} & \multicolumn{2}{c|}{1.958928121E-5} & \multicolumn{2}{c}{8.821066188E-8}\tabularnewline
\hline 
\end{tabular}
\end{table}

\begin{table}
\centering{}\caption{Exact solutions, numerical results, and relative errors for sensitivities
of moments at $\bm{\xi}_{0}=\left(0,0\right)$ - coarse mesh}
\label{Tab:complex-pressure-disk-mom-sen-coarse}%
\begin{tabular}{c|c|>{\centering}m{1.3cm}|c|>{\centering}m{1.3cm}|c|>{\centering}m{1.3cm}}
\hline 
\multirow{1}{*}{} & \multicolumn{2}{c|}{$D_{T}m^{(1)}(\Omega,\bm{\xi}_{0})$} & \multicolumn{2}{c|}{$D_{T}m^{(2)}(\Omega,\bm{\xi}_{0})$} & \multicolumn{2}{c}{$D_{T}m^{(3)}(\Omega,\bm{\xi}_{0})$}\tabularnewline
\cline{2-7} \cline{3-7} \cline{4-7} \cline{5-7} \cline{6-7} \cline{7-7} 
 & values  & Relative Error (\%)  & values  & Relative Error (\%)  & values  & Relative Error (\%)\tabularnewline
\hline 
PDD $S=1,\ m=1$  & 2.17057140E-4  & 0.422  & 1.90394853E-6  & 1.800  & 1.26338374E-8  & 3.444\tabularnewline
PDD $S=1,\ m=2$  & 2.17096271E-4  & 0.404  & 1.90509883E-6  & 1.741  & 1.26550116E-8  & 3.282\tabularnewline
PDD $S=1,\ m=3$  & 2.17074018E-4  & 0.414  & 1.90489284E-6  & 1.752  & 1.26538907E-8  & 3.291\tabularnewline
PDD $S=2,\ m=1$  & 2.17834040E-4  & 0.066  & 1.91224312E-6  & 1.372  & 1.27006070E-8  & 2.934\tabularnewline
PDD $S=2,\ m=2$  & 2.17282453E-4  & 0.319  & 1.90734319E-6  & 1.625  & 1.26761294E-8  & 3.121\tabularnewline
PDD $S=2,\ m=3$  & 2.18166027E-4  & 0.087  & 1.91619922E-6  & 1.168  & 1.27424908E-8  & 2.614\tabularnewline
\hline 
Exact  & \multicolumn{2}{c|}{2.179771038E-4} & \multicolumn{2}{c|}{1.938851314E-6} & \multicolumn{2}{c}{1.308450116E-8}\tabularnewline
\hline 
\end{tabular}
\end{table}

\begin{table}
\centering{}\caption{Exact solutions, numerical results, and relative errors for sensitivities
of moments at $\bm{\xi}_{0}=\left(0,0\right)$ - fine mesh}
\label{Tab:complex-pressure-disk-mom-sen-fine}%
\begin{tabular}{c|c|>{\centering}m{1.3cm}|c|>{\centering}m{1.3cm}|c|>{\centering}m{1.3cm}}
\hline 
\multirow{1}{*}{} & \multicolumn{2}{c|}{$D_{T}m^{(1)}(\Omega,\bm{\xi}_{0})$} & \multicolumn{2}{c|}{$D_{T}m^{(2)}(\Omega,\bm{\xi}_{0})$} & \multicolumn{2}{c}{$D_{T}m^{(3)}(\Omega,\bm{\xi}_{0})$}\tabularnewline
\cline{2-7} \cline{3-7} \cline{4-7} \cline{5-7} \cline{6-7} \cline{7-7} 
 & values  & Relative Error (\%)  & values  & Relative Error (\%)  & values  & Relative Error (\%)\tabularnewline
\hline 
PDD $S=1,\ m=1$  & 2.17130315E-4  & 0.389  & 1.91834626E-6  & 1.058  & 1.28213357E-8  & 2.011\tabularnewline
PDD $S=1,\ m=2$  & 2.17156561E-4  & 0.376  & 1.91936534E-6  & 1.005  & 1.28417041E-8  & 1.856\tabularnewline
PDD $S=1,\ m=3$  & 2.17147692E-4  & 0.381  & 1.91930875E-6  & 1.008  & 1.28418034E-8  & 1.855\tabularnewline
PDD $S=2,\ m=1$  & 2.17245778E-4  & 0.336  & 1.91970278E-6  & 0.988  & 1.28347047E-8  & 1.909\tabularnewline
PDD $S=2,\ m=2$  & 2.17177686E-4  & 0.367  & 1.91993928E-6  & 0.975  & 1.28503641E-8  & 1.789\tabularnewline
PDD $S=2,\ m=3$  & 2.17557373E-4  & 0.193  & 1.92318241E-6  & 0.808  & 1.28716567E-8  & 1.627\tabularnewline
\hline 
Exact  & \multicolumn{2}{c|}{2.179771038E-4} & \multicolumn{2}{c|}{1.938851314E-6} & \multicolumn{2}{c}{1.308450116E-8}\tabularnewline
\hline 
\end{tabular}
\end{table}

\begin{figure}
\centering{}\includegraphics[scale=0.5]{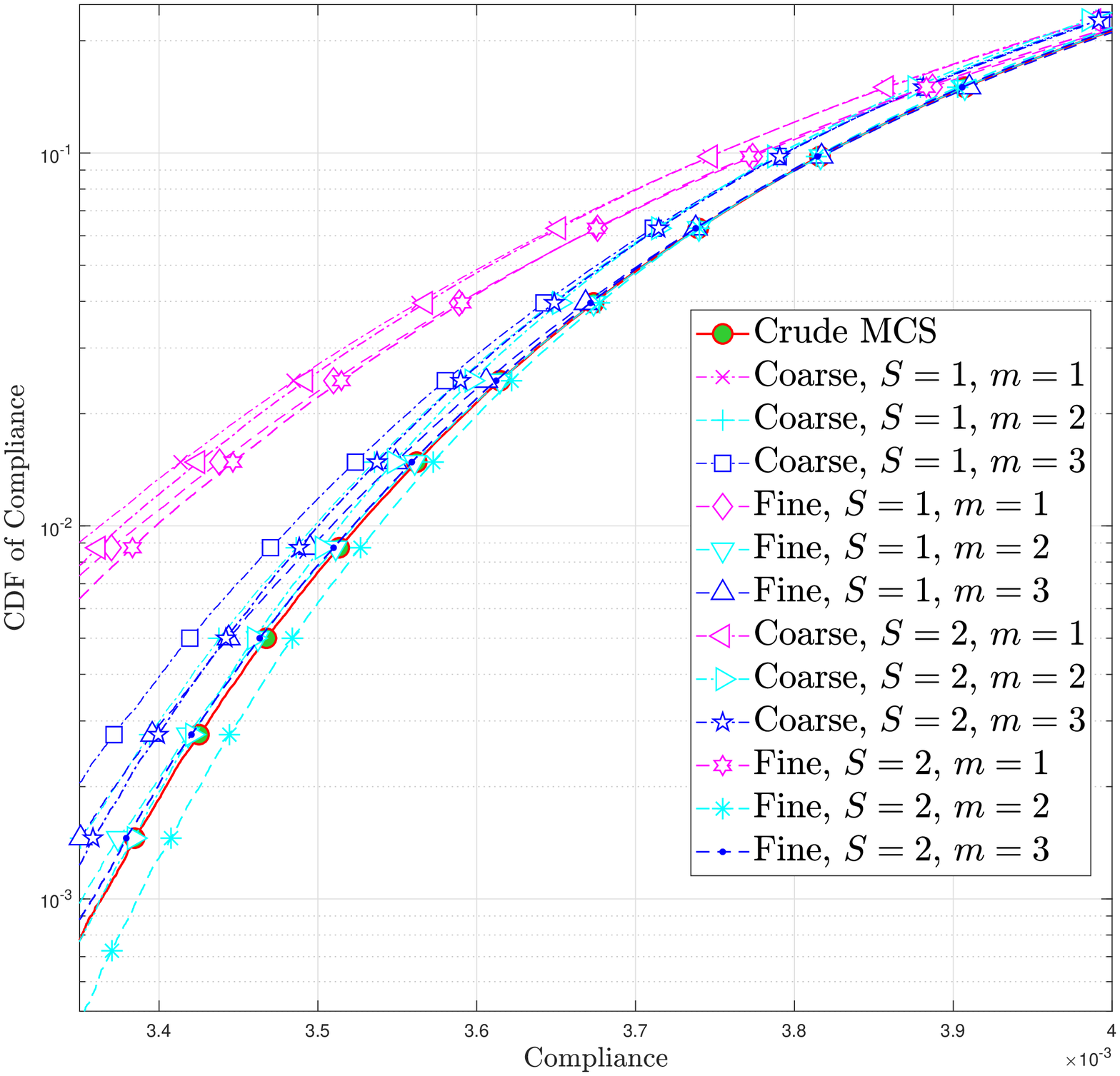}\caption{CDF of the compliance}
\label{Fig:CDF-compliance-disk-complex-pressure} 
\end{figure}

\begin{table}
\begin{centering}
\caption{Benchmark of reliability and its sensitivity for $\rho=0.05,\bar{y}=0.0036$
at $\bm{\xi}_{0}=\left(0,0\right)$ - coarse mesh}
\label{Tab:complex-pressure-disk-PF-sen-coarse}%
\begin{tabular}{c|c|c|c|c|c}
\hline 
\multirow{1}{*}{} & \multicolumn{2}{c|}{$P_{F}:=P\left(y\le\bar{y}\right)$, $\bar{y}=0.0036$} & \multicolumn{2}{c|}{$D_{T}P_{F}(\Omega,\bm{\xi}_{0})$} & \multirow{2}{*}{\# of FEA}\tabularnewline
\cline{2-5} \cline{3-5} \cline{4-5} \cline{5-5} 
 & values  & Relative Error (\%)  & values  & Relative Error (\%)  & \tabularnewline
\hline 
PDD $S=1,\ m=1$$\dagger$  & 4.86690000E-2  & 127.015  & -4.20000000E-2  & 15.170  & 107\tabularnewline
PDD $S=1,\ m=2$  & 2.70290000E-2  & 26.076  & -4.12000000E-2  & 12.976  & 107\tabularnewline
PDD $S=1,\ m=3$  & 3.02560000E-2  & 41.128  & -3.88000000E-2  & 6.395  & 213\tabularnewline
PDD $S=2,\ m=1$  & 4.77570000E-2  & 122.761  & -5.28000000E-2  & 44.785  & 5619\tabularnewline
PDD $S=2,\ m=2$  & 2.55110000E-2  & 18.995  & -4.92000000E-2  & 34.912  & 11131\tabularnewline
PDD $S=2,\ m=3$  & 2.79250000E-2  & 30.255  & -3.96000000E-2  & 8.588  & 22261\tabularnewline
\hline 
Crude MCS-FD$\ddagger$  & \multicolumn{2}{c|}{2.143872200E-2} & \multicolumn{2}{c|}{-3.646800000E-02} & NA\tabularnewline
\hline 
\end{tabular}
\par\end{centering}
\begin{centering}
$\dagger$ The sample size for results by proposed method is $L=10^{6}$ 
\par\end{centering}
\centering{}$\ddagger$ The sample size for the Crude MCS-FD is
$L=10^{9}$ 
\end{table}

\begin{table}
\begin{centering}
\caption{Benchmark of reliability and its sensitivity for $\rho=0.05,\bar{y}=0.0036$
at $\bm{\xi}_{0}=\left(0,0\right)$ - fine mesh}
\label{Tab:complex-pressure-disk-PF-sen-fine}%
\begin{tabular}{l|c|c|c|c|c}
\hline 
\multirow{1}{*}{} & \multicolumn{2}{c|}{$P_{F}:=P\left(y\le\bar{y}\right)$, $\bar{y}=0.0036$} & \multicolumn{2}{c|}{$D_{T}P_{F}(\Omega,\bm{\xi}_{0})$} & \multirow{2}{*}{\# of FEA}\tabularnewline
\cline{2-5} \cline{3-5} \cline{4-5} \cline{5-5} 
 & values  & Relative Error (\%)  & values  & Relative Error (\%)  & \tabularnewline
\hline 
PDD $S=1,\ m=1$$\dagger$  & 4.21980000E-2  & 96.831  & -4.20000000E-2  & 15.170  & 107\tabularnewline
PDD $S=1,\ m=2$  & 2.17680000E-2  & 1.536  & -3.48000000E-2  & 4.574  & 107\tabularnewline
PDD $S=1,\ m=3$  & 2.45970000E-2  & 14.732  & -3.36000000E-2  & 7.864  & 213\tabularnewline
PDD $S=2,\ m=1$  & 4.18240000E-2  & 95.086  & -4.72000000E-2  & 29.429  & 5619\tabularnewline
PDD $S=2,\ m=2$  & 1.99170000E-2  & 7.098  & -3.76000000E-2  & 3.104  & 11131\tabularnewline
PDD $S=2,\ m=3$  & 2.30350000E-2  & 7.446  & -3.60000000E-2  & 1.283  & 22261\tabularnewline
\hline 
Crude MCS-FD$\ddagger$  & \multicolumn{2}{c|}{2.143872200E-2} & \multicolumn{2}{c|}{-3.646800000E-2} & NA\tabularnewline
\hline 
\end{tabular}
\par\end{centering}
\begin{centering}
$\dagger$ The sample size for results by the proposed method is $L=10^{6}$ 
\par\end{centering}
\centering{}$\ddagger$ The sample size for the Crude MCS-FD is
$L=10^{9}$ 
\end{table}

For failure probability and its topology sensitivity, analytical expressions
or exact values are not readily available for this example. For simplicity,
the crude MCS that employs the analytical compliance Eq. \eqref{eq:disk-solution}
and a sample size $L=10^{9}$ is taken as the benchmark solution of
failure probability. Meanwhile, a finite difference formulation embedded
the crude MCS (Crude MCS-FD) 
\begin{equation}
D_{T}P\left[\mathbf{X}\in\Omega_{F}\right]\cong{\normalcolor {\normalcolor \frac{1}{\rho^{n}}}\lim_{L\rightarrow\infty}}\frac{1}{L}\sum_{l=1}^{L}\left[I_{\Omega_{F,\rho}}(\mathbf{x}^{(l)})-I_{\Omega_{F}}(\mathbf{x}^{(l)})\right]\label{eq:FD-DT-PF-1}
\end{equation}
is adopted as the benchmark solution of topology sensitivity of failure
probability, where the radius of the perforated hole takes a finite
value $\rho=0.05$, the sample size $L=10^{9}$, $I_{\Omega_{F}}$
and $I_{\Omega_{F,\rho}}$ are the indicator functions of the exact
failure domains $\Omega_{F}:=\{\mathbf{x}:y(\Omega,\mathbf{x})<\bar{y}\}$
and $\Omega_{F,\rho}:=\{\mathbf{x}:y(\Omega_{\rho},\mathbf{x})<\bar{y}\}$
with $y(\Omega,\mathbf{x})$ taking the exact compliance function
of the disk as shown in Eq. \eqref{eq:disk-solution} and $y(\Omega_{\rho},\mathbf{x})$
taking the exact compliance function of the perforated disk as shown
in Eq. \eqref{eq:ring-solution}. These benchmark solutions, involving
analytical expressions of compliance, MCS, and the finite-difference
method, is also referred to as semi-analytical solutions in this paper.

The cumulative distribution function (CDF) of the compliance by crude
MCS as well as ones by the proposed method employing two mesh cases
and various PDD truncations are plotted in Fig. \ref{Fig:CDF-compliance-disk-complex-pressure}.
An identical sample size $L=10^{6}$ is used for all plots in this
figure. All the CDF curves spontaneously group into two bundles. The
first bundle consists of all linear ($m=1$) approximations whether
univariate ($S=1$) or bivariate ($S=2$), fine mesh or coarse mesh.
It has considerable errors when comparing with the CDF of crude MCS,
indicating that the error due to lack of nonlinearity in the PDD dominants
the error from FEA and interactions between random variables. The
second bundle includes all the cases of $m\ge2$ and provide better
approximations. Among them, the cases using fine mesh provide better
solutions than coarse mesh ones. The best results are achieved by
two fine mesh cases - $S=1,m=2$ and $S=2,m=3$, and their curves
are almost coincide with the one by the crude MCS. Nonetheless, an
overall trend of convergence can be roughly observed in Fig. \ref{Fig:CDF-compliance-disk-complex-pressure}
as increasing $S$ and $m$ and adopting finer mesh. More quantitative
verifications of failure probability and its topology sensitivity
are displayed in Tables \ref{Tab:complex-pressure-disk-PF-sen-coarse}
and \ref{Tab:complex-pressure-disk-PF-sen-fine}, in which the failure
probability at $0.0036$ and its topology sensitivity are evaluated
by the proposed method and the crude MCS. The failure probability
by the linear approximations ($m=1$) carries the largest errors among
their same-variate and same-mesh counterparts, specifically $127.015$
and $122.761$ percent for coarse mesh $S=1,2$, $96.831$ and $95.086$
percent for fine mesh $S=1,2$. After increasing $m$, the errors
plummet dramatically to about $19$-$41$ percent for coarse mesh
cases and $2$-$14$ percent for fine mesh cases. The significant
differences in error levels of two mesh types imply that the error
from FEA predominates in those cases. Similar behaviors are observed
in the results of its topology sensitivity but the level of errors
have slight or moderate drops for most of $m\ge2$ cases. The proposed
method with the fine mesh and nonlinearity ($m\ge2$) provides satisfactory
evaluation for the topology sensitivity of failure probability, merely
$5$-$8$ percent for univariate and $1$-$3$ percent for bivariate
as shown in Table \ref{Tab:complex-pressure-disk-PF-sen-fine}. For
both failure probability and its sensitivity, Table \ref{Tab:complex-pressure-disk-PF-sen-coarse}-\ref{Tab:complex-pressure-disk-PF-sen-fine}
show that the error level roughly drops when increasing $S$ and $m$,
but the trend is not monotonic because of the synthetic effect of
four kinds of error sources - finite difference, MCS, PDD, and FEA.
The number of FEAs required by the proposed method for each PDD truncation
is also listed in Table \ref{Tab:complex-pressure-disk-PF-sen-coarse}-\ref{Tab:complex-pressure-disk-PF-sen-fine}.
Univariate cases are much more efficient than bivariate ones as expected,
involving only $107$ and $213$ FEAs to level down the errors to
$1.536$ and $14.732$ percent in failure probability and $4.574$
and $7.864$ in its topology sensitivity for fine mesh and $m=2,3$.
It is noteworthy that the same set of FEAs can be used to generate
estimations for not only failure probability and its sensitivity but
also moments and their sensitivity in preceding tables.

To sum up, this example is constructed to gauge the accuracy of new
or existing methods for stochastic analyses and their topology sensitivities
by analytical or semi-analytical solutions developed. Although $K=25$
is specified, the analytical and semi-analytical solutions developed
can be directly used or easily expanded for any positive $K$ to accommodate
even more random variables. Nonetheless, the proposed method is capable
of evaluating moments and their sensitivities in a highly accurate
manner even using low-variate low-order approximation. For failure
probability and its sensitivity, it is also feasible to provide satisfactory
evaluations using low-variate but nonlinear approximation. The least
number of FEAs required for those fine approximations is $107$ for
this 53 random variable example, demonstrating the high efficiency
of the proposed method for high-dimensional stochastic topology sensitivity
analysis. Another advantage of the proposed method observed in this
example is its capability of providing higher accuracy in topology
sensitivity than in stochastic quantities themselves.

\subsection{An engineering bracket involving 11 random variables}

\begin{figure}
\centering{}\includegraphics[scale=0.3]{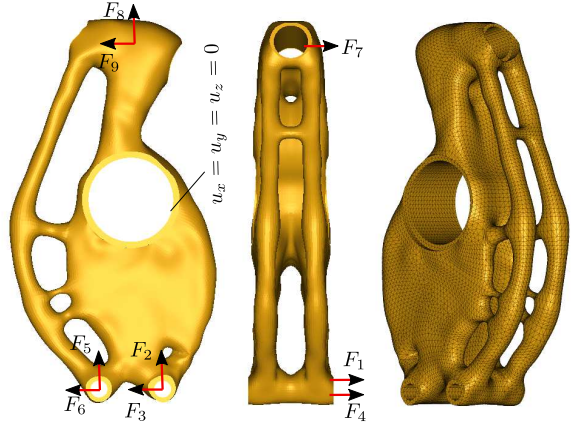}\caption{Geometry and mesh of the bracket}
\label{Fig:geo-mesh-alien-bracket} 
\end{figure}

Last, the proposed method is applied to a three-dimensional engineering
bracket \citep{grabcad} shown in Fig. \ref{Fig:geo-mesh-alien-bracket}.
With the fixed support at the middle hole, the bracket is subject
to nine random tractions along $x,\ y,\ $or $z$-direction on the
surfaces of one top hole and two bottom holes as shown in Fig. \ref{Fig:geo-mesh-alien-bracket}.
Their mean values are $\left(\mu_{F_{1}},\mu_{F_{2}},\cdots,\mu_{F_{9}}\right)=\left(2500.0,4200.0,-6400.0,3600.0,-5000.0,-6000.0,-4800.0,8100.0,-7000.0\right),$respectively.
The Young's modulus and Poisson's ratio are also random with mean
values $\mu_{E}=2.1\times10^{9}$ and $\mu_{\nu}=0.3$. The CV for
all 11 random variables is $0.1$. In this example, all 11 random
variables follow truncated Gaussian distribution, which has the following
PDF in general 
\begin{equation}
f_{X}\left(x\right)=\begin{cases}
\frac{1}{\Phi\left(D\right)-\Phi(-D)}\phi\left(\frac{x-\mu}{\sigma}\right) & \alpha\le x\le\beta,\\
0 & \mathrm{otherwise,}
\end{cases}
\end{equation}
where $\mu$ and $\sigma$ denote the mean and standard deviation
of each variable before the truncation and $\alpha=\mu-D,\ \beta=\mu+D$.
For nine random tractions and Young's modulus, $D$ takes 10 times
of the corresponding standard deviation, that is, $D=10\sigma$. For
the random Poisson's ratio, $D$ takes six times of the corresponding
standard deviation to avoid unrealistic materials.

\begin{figure}
\centering{}\includegraphics[scale=0.3]{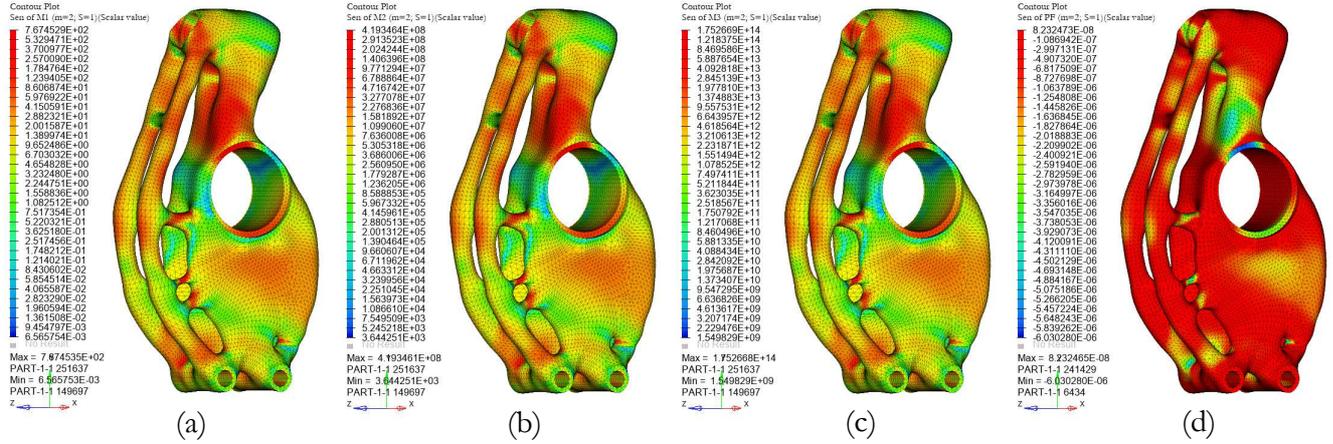}\caption{Stochastic topology sensitivity of compliance: (a)-(c) topology sensitivity
of 1st moment, 2nd moments, and 3rd moments; (d) topology sensitivity
of failure probability}
\label{Fig:contour-alien-bracket-sen-m3-pf} 
\end{figure}

The second-order univariate PDD ($S=1,m=2$) is used to perform stochastic
topology sensitivity analysis. The finite element model required contains
182540 quadratic tetrahedron elements. Compliance is selected as the
performance function $y$ and failure criteria for the reliability
is defined as $P_{F}:=P\left(y<1.6\times10^{5}\right)$. Contours
of stochastic topology sensitivities for compliance are plotted in
Fig. \ref{Fig:contour-alien-bracket-sen-m3-pf}. The contours for
sensitivities of the three moments follow similar patterns but different
value ranges as expected since the sensitivity is eventually related
to the stress field. The contour for the sensitivity of failure probability
is also similar due to the same reason although distinct colors manifest
the value difference. Only 23 FEAs are needed to evaluate the first
three moments, probability of failure, and their sensitivities for
this 11-dimensional example, illustrating the effectiveness of the
proposed method for high-dimensional engineering problems.

\section{Conclusions\label{sec7}}

A new framework for stochastic topology sensitivity analysis was developed
for solving RTO and RBTO problems commonly encountered in engineering.
The framework is grounded on the polynomial dimensional decomposition
and the concept of topology derivative. Comparing with previous developments,
the new method is capable of providing accurate evaluations of stochastic
topology sensitivity owing to the dovetailed topology derivative concept.
Furthermore, the new method can efficiently tackle high-dimensional
stochastic response functions and their topology sensitivities as
a result of the hierarchical structure of PDD which decomposes a high-dimensional
function in terms of lower-variate component functions. With these
two intrinsic advantages, the new method endows the first three moments
and their topology sensitivities with analytical expressions. And
it also provides embedded MCS for reliability analysis and finite
difference formulations for topology sensitivity of reliability. In
the finite difference formulations, the definition of topology derivative
is utilized as a callback to evaluate the perturbed performance function
requiring no additional function evaluations and thus results in a
self-consistent framework. It is noteworthy that the evaluation of
moments, reliability, and their topology sensitivity is acquired from
a sing stochastic analysis. In addition, the adjoint method inherited
from deterministic topology sensitivity analysis, together with PDD,
grant the proposed framework a significantly high efficiency for solving
high-dimensional engineering problems especially when FEA is involved.

Two new benchmark examples were developed to address the issue of
lacking analytical solutions of stochastic topology sensitivity for
verification. The first example provides not only the analytical expression
for the first three moments of compliance and their topology sensitivities
but also the analytical expression for the failure probability and
its topology sensitivity. Aided by this example, the accuracy and
efficiency of the proposed method are examined and demonstrated. The
second example, accommodating 53 random variables via applying an
intricate pressure, supplies analytical solutions for compliance of
both the original domain and the perforated domain. These analytical
compliances generate exact solutions for the moments and their sensitivities,
and also offer a precise evaluation of failure probability via crude
Monte Carlo simulation as well as an accurate assessment for its topology
sensitivity by virtue of finite difference method. The effectiveness
of the proposed method is thus verified and the advantages of the
dovetailed decomposition are illustrated by this 53-dimension example.
It also demonstrates that topology sensitivities of moments by the
proposed method possess higher accuracies than moments themselves
when the function structure of deterministic topology derivative is
simpler than the response itself. A similar advantage is also observed
in the topology sensitivity of failure probability in this example.
The proposed method is finally applied to a three-dimension bracket
with 11 random variables, by which the application to complex engineering
problems is examined.

In summary, the introduction of the topology derivative concept enables
a rigorous description of stochastic topology sensitivity and permits
the development of new benchmark examples for this research field.
The grounded polynomial dimensional decomposition empowers its high
efficiency to solve stochastic topology sensitivity for high-dimensional
complex engineering problems. In addition, when the deterministic
topology derivative of response takes a simpler form than the response
itself, the proposed method often supplies better accuracies on stochastic
topology sensitivities than on the stochastic analysis. 

\section*{Acknowledgments}

The authors acknowledge financial support from the U.S. National Science
Foundation under Grant No. CMMI-1635167 and the startup funding of
Georgia Southern University. Also to commemorate Niels Henrik Abel.\\

\appendix
%dummy comment inserted by tex2lyx to ensure that this paragraph is not empty

\section{The solutions for the external problem}

\subsection*{\label{appendix1}}

The solution of Eq. \eqref{eq:BVP-external} were well studied by
mathematicians in early research \citep{little1973elasticity,garreau2001topological}.
However, topological derivatives require only the solution on the
boundary $\partial\omega_{\rho}$, which can be obtained in an easier
approach comparing to those in literature \citep{little1973elasticity,garreau2001topological}.
In this appendix, an approach based on Eshelby tensor \citep{eshelby1975elastic}
and solutions for plane stress, plane strain, and three-dimensional
cases are compiled for easy accessibility of researchers in mechanics
and engineering field. When the elastic medium in Eshelby phase-transition
strain problem is isotropic and the inclusion domain $\Omega$ is
a sphere, the Eshelby tensor is isotropic 
\begin{equation}
\mathbb{S}=\left(\alpha-\beta\right)\frac{1}{3}\boldsymbol{\delta\delta}+\beta\mathbb{I}
\end{equation}
where

\[
\alpha=\frac{3K}{3K+4G},\ \ \ \beta=\frac{6\left(K+2G\right)}{5\left(3K+4G\right)},
\]
and $G$ and $K$ are shear modulus and bulk modulus, respectively.
The real strain on the boundary of the inclusion reads 
\begin{equation}
\hat{\boldsymbol{\epsilon}}=\left(\mathbb{S}^{-1}-\mathbb{I}\right)^{-1}\mathbb{C}^{-1}:\hat{\boldsymbol{\sigma}}
\end{equation}
where $\hat{\boldsymbol{\sigma}}$ is the stress on the surface of
the inclusion. To utilize it for the solution on $\partial\omega_{\rho}$
of Eq. \eqref{eq:BVP-external}, let

\begin{equation}
\hat{\boldsymbol{\sigma}}=\boldsymbol{\sigma}\left(\boldsymbol{\xi}_{0}\right)
\end{equation}
where $\boldsymbol{\sigma}\left(\boldsymbol{\xi}_{0}\right)$ is the
stress at $\boldsymbol{\xi}_{0}$ in Eq. \eqref{eq:BVP}. Therefore
the strain solution for Eq. \eqref{eq:BVP-external}

\begin{equation}
\hat{\boldsymbol{\epsilon}}=\left(\frac{a-b}{3}\boldsymbol{\delta\delta}+b\mathbb{I}\right):\boldsymbol{\sigma}\left(\boldsymbol{\xi}_{0}\right)
\end{equation}
where $a=\frac{1}{4G}=\frac{1+\nu}{2E},\ b=\frac{3\left(K+2G\right)}{G\left(9K+8G\right)}=\frac{2(4-5\nu^{2}-\nu)}{E(7-5\nu)}$.
The corresponding displacement solution on $\partial\omega_{\rho}$
reads 
\begin{align}
\hat{\boldsymbol{u}} & =\left(\frac{a-b}{3}\boldsymbol{\delta\delta}+b\mathbb{I}\right):\boldsymbol{\sigma}\left(\boldsymbol{\xi}_{0}\right)\cdot\boldsymbol{n}\rho\nonumber \\
 & =\rho\left(\frac{a-b}{3}\mathrm{tr}\left(\boldsymbol{\sigma}\left(\boldsymbol{\xi}_{0}\right)\right)\boldsymbol{n}+b\boldsymbol{n}\cdot\boldsymbol{\sigma}\left(\boldsymbol{\xi}_{0}\right)\right)\label{eq:del_u using Eshelby}
\end{align}

For plane strain cases, the Eshelby tensor becomes 
\begin{equation}
\mathbb{S}=\left(\alpha-\beta\right)\frac{1}{2}\boldsymbol{\delta\delta}+\beta\mathbb{I}
\end{equation}
with $\alpha=\frac{1}{2(1-\nu)},\ \ \beta=\frac{3-4\nu}{4(1-\nu)}$,
and the displacement solution on $\partial\omega_{\rho}$ becomes
\begin{align}
\hat{\boldsymbol{u}} & =\frac{\left(1+\nu\right)}{E}\left[\left(2\nu-1\right)\boldsymbol{\delta\delta}+\left(3-4\nu\right)\mathbb{I}\right]:\boldsymbol{\sigma}\left(\boldsymbol{\xi}_{0}\right)\cdot\boldsymbol{n}\rho\nonumber \\
 & =\rho\frac{\left(1+\nu\right)}{E}\left[\left(2\nu-1\right)\mathrm{tr}\left(\boldsymbol{\sigma}\left(\boldsymbol{\xi}_{0}\right)\right)\boldsymbol{n}+\left(3-4\nu\right)\boldsymbol{n}\cdot\boldsymbol{\sigma}\left(\boldsymbol{\xi}_{0}\right)\right].
\end{align}
For plane stress cases, simply changing the elastic constant, we have
\begin{align}
\hat{\boldsymbol{u}} & =\rho\left[\frac{\nu-1}{E}\mathrm{tr}\left(\boldsymbol{\sigma}\left(\boldsymbol{\xi}_{0}\right)\right)\boldsymbol{n}+\frac{3-\nu}{E}\boldsymbol{n}\cdot\boldsymbol{\sigma}\left(\boldsymbol{\xi}_{0}\right)\right]
\end{align}

\noindent \bibliographystyle{elsarticle-num}
\bibliography{references/thesis,references/top-derivative,references/ref-my-pub,references/utility,references/RBTO-single-loop,references/RBTO-double-loop,references/f-div,references/level-set2,references/density,references/PCE,references/RBTO,plbib,references/RTO-RBTO-new,references/elasticity-solid-mechanics,references/Misc}

\end{document}